\documentclass[twoside,11pt]{article}
\renewcommand{\baselinestretch}{1.7}

\usepackage{graphicx}
\usepackage{graphics,ifthen}
\usepackage{latexsym}
\usepackage{amsfonts}
\usepackage{amsmath}
\usepackage{amssymb}%
\usepackage{amsthm}
\usepackage{latexsym}
\usepackage{amscd}
\begin{document}
\input{latexP.sty}
\input{referencesP.sty}
\input{epsf.sty}

\def\e{\text{\hskip1.5pt e}}
\newcommand{\eps}{\epsilon}
\newcommand{\remarks}{\noindent {\bf Remarks:\ }}
\newcommand{\note}{\noindent {\bf Note:\ }}
\newcommand{\Lower}[2]{\smash{\lower #1 \hbox{#2}}}

\newtheorem{thm}{Theorem}[section]
\newtheorem{examp}{Example}[section]
\newtheorem{defin}{Definition}[section]
\newtheorem{prop}{Proposition}[section]
\newtheorem{lem}{Lemma}[section]
\newtheorem{cor}{Corollary}[section]
\newtheorem{rem}{Remark}[section]
\newtheorem{algorithm}{Algorithm}[section]

\def\I{\mathbb{I}}
\def\E{\mathbb{E}}
\def\C{\mathbb{C}}
\def\V{\mathbb{V}}
\def\P{\mathbb{P}}
\def\cU{\cal{U}}
\def\T{{\bf T}}
\def\J{{\bf J}}
\def\Q{{\bf Q}}
\def\R{\cal{R}}
\def\Rp{\cal{H}}
\def\p{{\bf p}}
\def\S{{\bf S}}
\def\X{{\bf X}}
\def\u{{\bf u}}
\def\vv{{\bf v}}
\def\y{{\bf y}}
\def\z{{\bf z}}
\def\Z{{\bf Z}}
\def\U{{\bf U}}
\def\Y{{\bf Y}}
\def\e{e}
\def\Np{N(\p)}
\def\np{n(\p)}
\def\fN{f_{\Lower{0.5ex}{{\scriptsize $N$}}}}
\def\ftN{f_{\Lower{0.25ex}{{\tiny $N$}}}}
\def\gN{g_{\Lower{0.5ex}{{\scriptsize $N$}}}}
\def\gtN{g_{\Lower{0.25ex}{{\tiny $N$}}}}
\def\muN{\mu_{\Lower{0.5ex}{{\scriptsize $N$}}}}
\def\mugN{\mu(\gN)}
\def\mugtN{\mu(\gtN)}
\def\Bt{\mbox{\boldmath{$\beta$}}}
\def\fNtt{f_{\Lower{0.5ex}{{\scriptsize $N$,{\scriptsize$\theta$}}}}}
\def\ftNtt{f_{\Lower{0.25ex}{{\tiny $N$,{\tiny$\theta$}}}}}
\def\gNtt{g_{\Lower{0.5ex}{{\scriptsize $N$,{\scriptsize$\theta$}}}}}
\def\gtNtt{g_{\Lower{0.25ex}{{\tiny $N$,{\tiny$\theta$}}}}}
\def\muNtt{\mu_{\Lower{0.5ex}{{\scriptsize $N$,{\scriptsize$\theta$}}}}}
\def\mugNtt{\mu(\gNtt)}
\def\mugtNtt{\mu(\gtNtt)}
\def\sBt{\mbox{\boldmath{\scriptsize$\beta$}}}
\def\fNt{f_{\Lower{0.5ex}{{\scriptsize $N$,\mbox{\boldmath{\scriptsize$\beta$}},\scriptsize $\t$}}}}
\def\ftNt{f_{\Lower{0.25ex}{{\tiny $N$,\mbox{\boldmath{\tiny$\beta$}},\tiny $\t$}}}}
\def\gNt{g_{\Lower{0.5ex}{{\scriptsize $N$,\mbox{\boldmath{\scriptsize$\beta$}},\scriptsize $\t$}}}}
\def\gtNt{g_{\Lower{0.25ex}{{\tiny $N$,\mbox{\boldmath{\tiny$\beta$}},\tiny $\t$}}}}
\def\muNt{\mu_{\Lower{0.5ex}{{\scriptsize $N$,\mbox{\boldmath{\scriptsize$\beta$}},\scriptsize $\t$}}}}
\def\mugNt{\mu(\gNt)}
\def\mugtNt{\mu(\gtNt)}

\def\mugtNt0{\mu(g_{\Lower{0.25ex}{{\tiny $N$,{\tiny$0$}}}})}

\def\pst{\mbox{\boldmath$\bf\nu$}}               
\def\tpst{\mbox{\scriptsize \boldmath${\bf \nu}$}}  
\def\ps{\mbox{\boldmath${\bf \xi}$}}                
\def\tps{\mbox{\scriptsize \boldmath${\bf \xi}$}}   
\def\ftps{\mbox{\tiny \boldmath${\bf \xi}$}}        
\def\ftpst{\mbox{\tiny \boldmath${\bf \nu}$}}        
\def\Eps{E_{\mbox{\scriptsize \boldmath${\bf \xi}$}}}   
\def\Etps{E_{\mbox{\tiny \boldmath${\bf \xi}$}}}        
\def\CP{\mathcal{P}}
\def\CM{\mathcal{M}}

\def\Report{Bathtub hazard}
\def\Author{Ho}
\pagestyle{myheadings} \markboth{\Author}{\Report}
\thispagestyle{empty} \bct\Heading A Bayes method for a Bathtub
Failure Rate via two $\S$-paths \lbk\lbk \smc Man-Wai Ho\footnote{\rm Man-Wai Ho
is Assistant Professor, Department of Statistics and
Applied Probability, National University of Singapore, 6
Science Drive 2, Singapore 117546 (E-mail:
\textit{stahmw\at nus.edu.sg}). This work was partially
supported by National University of Singapore research grant 
R-155-050-067-131 and R-155-050-067-101.} \lbk\lbk
\BigSlant National University of Singapore\rm \lbk
(\today) \ect 
\begin{abstract}
A class of semi-parametric hazard/failure rates with a
bathtub shape is of interest. It does not only provide a
great deal of flexibility over existing parametric methods
in the modeling aspect but also results in a closed and
tractable Bayes estimator for the bathtub-shaped failure
rate~(BFR). Such an estimator is derived to be a finite sum
over two $\S$-paths due to an explicit posterior analysis
in terms of two~(conditionally independent) $\S$-paths.
These, newly discovered, explicit results can be proved to
be a Rao-Blackwellization of counterpart results in terms
of partitions that are readily available by a
specialization of James~(2005)'s work. We develop both
iterative and non-iterative computational procedures based
on existing efficient Monte Carlo methods for sampling one
single $\S$-path. Numerical simulations are given to
demonstrate the practicality and the effectiveness of our
methodology. Last but not least, two applications of the
proposed method are discussed, of which one is about a
Bayesian test for failure rates and the other is related to
modeling with covariates.
\end{abstract}
\rm

\section{Introduction}
In reliability theory and survival analysis it is often
important to understand a hazard rate~(or failure rate) as
it is interpreted as the propensity of failure of an item
or death of a human being in the instant future given its
survival until time $t$. There are a variety of shapes for
the function, for example, constant, non-increasing, or
non-decreasing, of which each corresponds to a different
life distribution. In particular, a class of life
distributions which corresponds to a bathtub-shaped failure
rate~(BFR) has received considerable attention as most
electronic, eletromechanical, and mechanical products and
human beings are subject to a high risk for failures/deaths
initially in an ``infant mortality'' phase, then to a lower
and constant risk in the so-called ``useful life'' period
and finally to an increasing risk with time during the
so-called ``wearout'' phase. Many parametric families of
distributions for
BFRs have been proposed over the last few decades. 
Most of which typically involving three or more parameters
are based on mixtures or generalizations of some common
probability distributions, such as exponential, gamma,
Weibull and Pareto distributions; see Rajarshi and
Rajarshi~(1988) and Lai, Xie, and Murthy~(2001, Section 4)
for an extensive and collective review. For discussion of
parametric models for other typical hazard functions, see
Kalbfleisch and Prentice~(1980) and Lawless~(1982). Also
see Singpurwalla~(2006) for a comprehensive discussion on
reliability and risk from a Bayesian perspective.

One of the contributions of the present paper is a closed
and tractable nonparametric estimator of BFRs that serve as
a viable estimator of any BFR and, hence, an alternative to
most existing parametric inferences which suffer from
intractability problems~[Lawless~(1982),~Page 255] and
often resort to extensive iterative procedure~[Haupt and
Schabe~(1997)]. The literature on nonparametric estimation
of BFRs is rather limited though there are some available
testing procedures involving BFRs~(see, for example,
Bergman~(1979), Aarset~(1985) and Vaurio~(1999)).
Amman~(1984)~(see also Laud, Damien and Walker~(2006))
studied a $U$-shaped process by combining two random
processes, of which one is the increasing random hazard
rates based on extended gamma processes firstly considered
by Dykstra and Laud~(1981) and the other one is the
decreasing counterpart defined analogously. However, the
combined process does not necessarily generate BFRs.
Reboul~(2005) introduced a data-driven nonparametric
estimator of BFRs which, though is not in a closed form,
can be computed by applying the ``Pool Adjacent Violators
Algorithm''~(see Barlow, Bartholomew, Bremner, and
Brunk~(1972)). References on nonparametric inference of any
of hazard, survivor, or cumulative hazard functions in
survival analysis include, for instance, Kaplan and
Meier~(1958), Watson and Leadbetter~(1964a,b),
Nelson~(1969), Doksum~(1974), Susarla and Van Ryzin~(1976),
Aalen~(1978), Ferguson and Phadia~(1979), Tanner and
Wong~(1983), Yandell~(1983), Lo and Weng~(1989),
Hjort~(1990), Wolpert and Ickstadt~(1998) and James~(2005),
among others; see Ghosh and Ramamoorthi~(2003) for a review
of works related to Bayesian nonparametrics, and see also
Sinha and Dey~(1997) for an extensive survey on
semi-parametric modeling of survival data with presence of
covariates.

In line with James~(2005) who studied random hazard rates
with general shapes expressible as $ \lambda(x|\mu) = \int
K(x,u) \mu(du),$ wherein $K(x,u)$ is a known positive
measurable kernel on a Polish space $\mathcal{X}\times
\mathcal{U}$ and $\mu$ is a completely random
measure~[Kingman~(1967, 1993)] on $\mathcal{U}$~(see Lo and
Weng~(1989) for the case when $\mu$ is an extended/weighted
gamma random measure), the present paper considers a
semi-parametric family of hazard rates on ${\Rp}= ( 0,
\infty)$ defined by, for $t,\t\in {\Rp}$,
\begin{equation}\label{bathtub}
    \lambda(t|\mu,\t) = \int_{\R}
    [\I(t-\t \leq u < 0) + \I(0 < u \leq t-\t)]
    \mu(du),
\end{equation}
where $\I(A)$ is the indicator function of a set $A$ and
$\mu$ is a completely random measure on $\R=
(-\infty,\infty)$. Argument of Brunner~(1992) in
constructing unimodal densities on the real line with mode
$\t$ based on the mixture representation of a monotone
failure rate~(MFR) considered by Lo and Weng~(1989) applies
and justifies that~(\ref{bathtub}) gives an BFR on $\Rp$
with a minimum point, or a change point called by Mitra and
Basu~(1995), at $\t \in \Rp$. Posterior consistency of
these BFRs can be established following Dr\v{a}gichi and
Ramamoorthi~(2003) who showed the corresponding result for
the class of MFRs discussed in Ho~(2006a), a subclass
of~(\ref{bathtub}) when $\t = 0$ or $\t=\infty$. Exploiting
the fine structure of an indicator kernel, Ho~(2006a)
improves the readily available explicit posterior analysis
in terms of partitions in James~(2005, Section 4) by giving
a tractable and less complex~(see Brunner and Lo~(1989))
characterization in terms of one $\S$-path for such MFRs,
and shows that an efficiently designed algorithm for
sampling an $\S$-path, called the accelerated path~(AP)
sampler, results in less variable Bayes estimates of the
hazard compared to a partition-based algorithm introduced
by James~(2005) via numerical simulations. In this work, we
show that all BFRs defined in~(\ref{bathtub}) possess nice
and special structures that naturally arise in relation to
two conditionally independent $\S$-paths given $\t$ in
Section~\ref{sec:model}, rather than one in the case of
MFRs; for an BFR there are two~(possibly different)
non-decreasing curves away from the change point $\t$ in
either direction, compared with only one such curve to the
right of the origin for a non-decreasing hazard rate. In
particular, an explicit characterization depending on two
$\S$-paths possessed by all such BFRs, which are
unprecedentedly available, generalizes the corresponding
characterization of MFRs discussed in Ho~(2006a) that
depends on only one path, and, more importantly, yields a
tractable Bayes estimator of BFRs as a finite sum over two
$\S$-paths. Understanding these novel characterization and
estimator for BFRs is of statistical importance; they can
be shown to be a Rao-Blackwellization of the
partition-based counterparts, suggesting that more
parsimonious methods for inference, compared with
partition-based methods introduced in James~(2005), would
be available if one could efficiently sample the two paths
in this context. To approximate posterior quantities for
models in~(\ref{bathtub}), Section~\ref{sec:MC} proposes an
iterative Monte Carlo procedure based on the AP sampler.
Furthermore, extensions of a sequential importance
sampling~(SIS)~[Kong, Liu, and Wong~(1994) and Liu and
Chen~(1998)] scheme for sampling one path at a time are
introduced. Numerical results of the method are given in
Section~\ref{sec:sim} to demonstrate its practicality and
effectiveness. Two applications of the methodology are
given in the last two sections in which the proposed
algorithms can be applied to approximate the posterior
quantities of interest. A test of an MFR versus an BFR
based on models in~(\ref{bathtub}) is illustrated in
Section~\ref{sec:testing}. Section~\ref{sec:cox} shows that
a two $\S$-path characterization also exists in modeling
with covariates by a proportional hazards model.

\section{Posterior analysis via two $\S$-paths}\label{sec:model}
A class of random hazard rates with a bathtub shape on the
half line ${\Rp}$, defined by~(\ref{bathtub}), is of
interest. The law of $\mu$ is uniquely characterized by the
Laplace functional
\begin{equation}
\label{Lap}
    {\cal L}_{\mu}(g|\rho, \eta) = \exp\left[ -\int_{\R} \int_{\Rp}
    \left(1- \e^{-g(u)x}\right)
    \rho(dx|u) \eta(du) \right],
\end{equation}
where $g$ is a non-negative function on $\R$ and
$\rho(dx|u) \eta(du)$ is called the L\'{e}vy measure of
$\mu$. Also, $\mu$ can be represented in a distributional
sense as
$$
    \mu(du) = \int_{\Rp} x {\cal N}(dx, du),
$$
where ${\cal N}(dx, du)$ is a Poisson random measure,
taking on points $(x,u)$ in $\Rp \times \R$, with mean
intensity measure
\begin{equation}
\label{intensity}
    \E[{\cal N}(dx, du)] = \rho(dx|u) \eta(du),
\end{equation}
such that $\int_{B} \int_{\Rp} \mbox{min}(x,1) \rho(dx|u)
\eta(du) < \infty$ for any bounded set $B \in \R$.

Suppose we collect independent failure times $\T=(T_1,
\ldots, T_N)$ from $N$ items with a common continuous life
distribution which corresponds to an BFR with change point
at $\t$, specified by~(\ref{bathtub}), until time $\tau$,
so that $0 < T_{1} < \cdots < T_{m} < \tau$ denote $m$
completely observed failure times, and $T_{m+1}=\cdots
=T_{N} \equiv \tau $ are $N_c \equiv N-m$ number of
right-censored times. Assuming a multiplicative intensity
model discussed in Aalen~(1975, 1978), the likelihood of
the data $\T$ is proportional to
\begin{equation}
\label{like}
    \e^{-\mugtNtt}
    \prod_{i=1}^{m} 
    \int
    [\I(T_i-\t \leq u_i < 0) + \I(0 < u_i \leq T_i-\t)] \mu(du_i),
\end{equation}
where
$$
\gNtt(u) = \int_{0}^{\tau} \left[\sum_{i=1}^{N} \I(T_i \geq t)
    \right] [\I(t-\t \leq u < 0) + \I(0 < u \leq t-\t)] dt
$$
is a piecewise linear function of $u$, and $\mugNtt =
\int_{\R} \gNtt(u) \mu(du) = \int_0^{\tau}
\left[\sum_{i=1}^{N} \I(T_i \geq t)\right]
\lambda(t|\mu,\t) dt$ with $\sum_{i=1}^{N} \I(T_i \geq t)$
called the total time on test~(TTT) transform~[Barlow,
Bartholomew, Bremner, and Brunk~(1972)]. Define $\fNtt(x,u)
= \gNtt(u)x$ for any $(x,u) \in (\Rp,\R)$ and assume that
\begin{equation}
\label{kappa}
    \kappa_{\ell}(\e^{-\ftNtt} \rho|u) =
    \int_{\R} x^{\ell} \e^{-\gtNtt(u)x} \rho(dx|u) < \infty,
\end{equation}
for any positive integer $\ell \leq m$ and a fixed $u\in\R$.

The posterior distribution of the pair $(\mu,\theta)$
in~(\ref{bathtub}) given $\T$ with respect to any prior
$\pi(d\theta)$ for $\theta \in \Rp$ can always be
determined by the double expectation formula,
\begin{equation}\label{double}
\E[h(\mu,\theta) |\T ] = \E\{\E[h(\mu,\theta) |\theta,\T ]|\T\} =
\int_{\Rp} \int_{\CM} h(\mu,\theta) \CP(d\mu|\theta,\T)
\CP(d\theta|\T),
\end{equation}
where $h$ is any nonnegative or integrable function, $\CM$ is the
space of measures over $\R$, and, $\CP(d\mu|\theta,\T)$ and
$\CP(d\theta|\T)$ denote the conditional distribution of $\mu$ given
$(\theta,\T)$ and the posterior distribution of $\theta$ given $\T$,
respectively.

\begin{figure}[h]
  \includegraphics[width=5.5in]{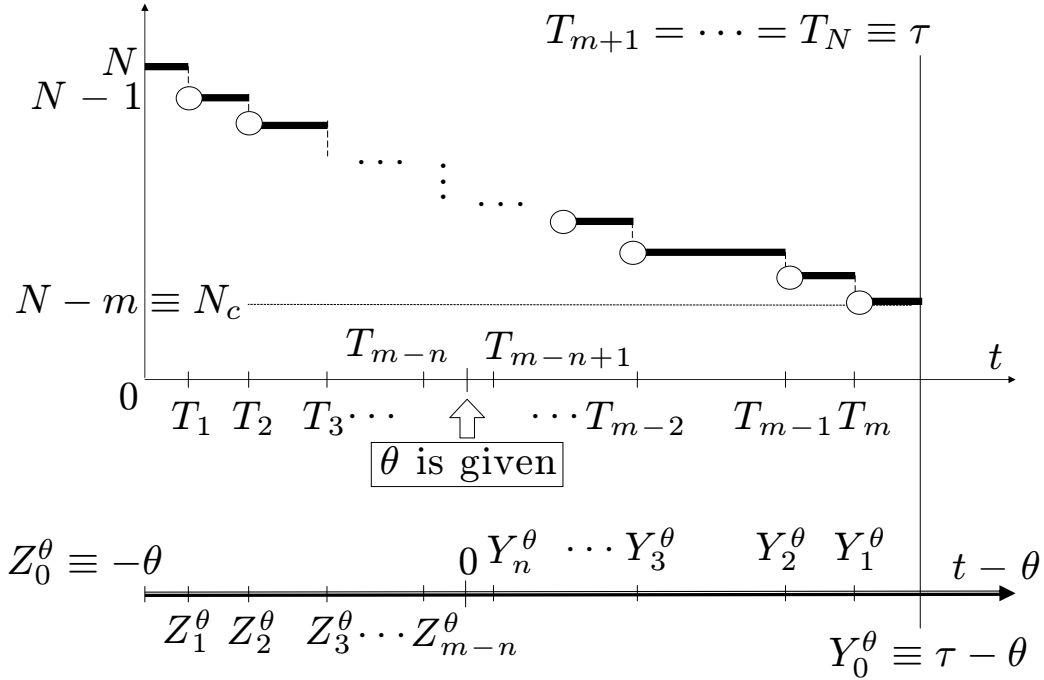}\\
\renewcommand{\baselinestretch}{1}
  \centering\parbox{5in}
  {\caption{\quad Illustration of the TTT transform and the relationship~(\ref{obs2}) between $\T$ and $(\Y^\t,\Z^\t,\t)$.}
  \label{fig:TTT}}
\end{figure}

Let us first look at $\CP(d\mu|\theta,\T)$ and then discuss
$\CP(d\theta|\T)$ later on. Suppose $0 < \t < \tau $, we
can always assume that
\begin{equation}\label{obs2}
(T_1-\theta,\ldots, T_m-\theta) = \Z^{\t} \cup \Y^{\t} =
(Z_{1}^{\t},Z_{2}^{\t}, \ldots,Z_{m-n}^{\t}) \cup
(Y_1^{\t},Y_2^{\t},\ldots,Y_{n}^{\t}),
\end{equation}
where $-\t \equiv Z^{\t}_{0} < Z_{1}^{\t}<Z_{2}^{\t} <
\cdots < Z_{m-n}^{\t}< Z^{\t}_{m-n+1} \equiv 0$ and $0
\equiv Y^{\t}_{n+1} <Y_n^{\t} < Y_{n-1}^{\t} < \cdots <
Y_{1}^{\t} < Y^{\t}_{0} \equiv \tau - \t$ are referred to
as negative and positive observations in the sequel. The
relationship between these notation and the data $\T$ is
illustrated in Figure~\ref{fig:TTT}, graphed together with
the TTT transform. It is worthy of note that once a failure
time $T_i$, $i=1,\ldots,m$, is completely observed and
compared with the given $\t$, the mixture hazard rates can
be simplified as in one of two mutually exclusive
situations specified by
\begin{equation}\label{twocases}
\lambda(T_i|\mu,\theta) = \left\{
\begin{array}{lll}
\int \I(Z_j^\t \leq u_i < 0) \mu(du_i), && T_i-\theta = Z_j^\t< 0, \\
\int \I(0 < u_i \leq Y_k^\t) \mu(du_i), && T_i-\theta=
Y_k^\t > 0,
\end{array}
\right.
\end{equation}
for $j=1,\ldots,m-n$ and $k=1,\ldots,n$. This also implies
that the missing variable $u_i$ corresponding to $T_i$
in~(\ref{like}) is always greater~(resp. smaller) than 0 if
$T_i>\hspace*{-0.06in}~(\mbox{resp. }<)\t$. This nice
similification proves to be crucial in leading to the
tractable path structure of BFRs in~(\ref{bathtub}).

Define an integer-valued vector $\S = (S_0, S_1, \ldots,
S_{m-1}, S_m)$~[Lo and Weng~(1989) and Brunner and
Lo~(1989)], referred to as an $\S$-path (of $m+1$
coordinates), which satisfies $S_0 = 0$, $S_m = m$ and $S_j
\leq \min(j,S_{j+1})$, $j=1,\ldots,m-1$. An $\S$-path is a
combinatorial reduction of a partition in the sense that an
$\S$-path of $m+1$ coordinates is said to
\textit{correspond to} one or many partitions
$\p=\{C_1,\ldots,C_{\np}\}$ of the integers
$\{1,\ldots,m\}$, provided that (i)~indices of the maximal
elements of the $\np$ cells $C_k$'s in $\p$ coincide with
locations $j$ at which $S_j
> S_{j-1}$, and (ii)~number of indices $e_k$ of cell $C_k$ for all
$k=1,\ldots,\np$ with a maximal index $j$, $j=1,\ldots,m$,
is identical to $S_j - S_{j-1}$. Given a path $\S$ of $m+1$
coordinates, let ${\mathbb{C}}_{\S}$ denote the collection
of all partitions that correspond to $\S$. Then, the total
number of partitions in ${\mathbb{C}}_{\S}$ is given
by~[Brunner and Lo~(1989)]
\begin{equation}\label{card}
|{\mathbb{C}}_{\S}| \equiv \sum_{\p \in {\C}_\S} 1 =
\displaystyle\prod_{\{j^{\ast}|\S\}} \binom{j - 1-S_{j-1}}{j-S_{j}
},
\end{equation}
where, conditioning on a path $\S$ of $m+1$ coordinates,
$\prod_{\{j^{\ast}|\S\}}$ stands for
$\prod_{j=1:S_j>S_{j-1}}^{m}$. Similarly,
$\sum_{\{j^{\ast}|\S\}}$ will stand for
$\sum_{j=1:S_j>S_{j-1}}^{m}$. See Ho~(2002) for more
discussion of the relationship between $\p$ and $\S$.

\begin{thm} \label{Prop1}
{\rm Suppose that the likelihood of the data $\T$ is given
by~(\ref{like}) and that $\mu$ is a completely random measure
characterized by the Laplace functional~(\ref{Lap}). Then, the
posterior distribution of $\mu$ given $\t$ and $\T$ can be described
as a mixture as follows:
\begin{itemize}
\item[(i)] Given $(\theta,\T)$, there are two paths
$\S^- = (0, S_1^-, \ldots,  S_{m-n-1}^-, m-n)$ and $\S^+ =
(0, S_1^+, \ldots, S_{n-1}^+, n)$, independently
distributed as
\begin{eqnarray}
    W^-( \S^-|\theta,\T) \propto \phi^-_\t(\S^-,\T) =
    |\mathbb{C}_{\S^-}| \prod_{\{j^{\ast}|\S^-\}}  \int_{Z^\t_j}^{0}
    \kappa_{m_j^-}(\e^{-\ftNtt} \rho|y)
    \eta(dy)\label{jointSn}\\
\noalign{\hspace*{0.37in}and}
    W^+( \S^+|\theta,\T) \propto \phi^+_\t(\S^+,\T) =
    |\mathbb{C}_{\S^+}| \prod_{\{j^{\ast}|\S^+\}} \int_{0}^{Y^\t_j}
    \kappa_{m_j^+}(\e^{-\ftNtt} \rho|y)
    \eta(dy),\label{jointSp}
\end{eqnarray}
where $|\mathbb{C}_{\S^-}|$ and $|\mathbb{C}_{\S^+}|$ are
defined in~(\ref{card}), $m^-_j\equiv
S^-_j-S^-_{j-1},j=1,\ldots,m-n$ and $m^+_j\equiv
S^+_j-S^+_{j-1},j=1,\ldots,n$.

\item[(ii)] Given $(\S^-,\S^+,\t,\T)$, there exist
$\sum_{\{j^{\ast}|\S^-\}}1$ and $\sum_{\{j^{\ast}|\S^+\}}1$
independent pairs of $(y_j^-, Q_j^-)$ and $(y_j^+, Q^+_j)$, denoted
by $(\y^-,\Q^-) = \{(y_j^-,Q_j^-): m_j^->0,j=1,\ldots,m-n\}$ and
$(\y^+,\Q^+) = \{(y_j^+,Q_j^+): m_j^+>0,j=1,\ldots,n\}$,
respectively. They are distributed as
\begin{eqnarray}
    \hspace*{-0.3in}\eta_j(dy_j^- |\S^-,\t,\T ) &\propto &\I(Z^\t_j \leq y_j^- < 0)
    \kappa_{m_j^-}(\e^{-\ftNtt} \rho|y_j^-)
    \eta(dy_j^-),\label{alphaSjn}\\
    \hspace*{-0.3in}\Pr\{Q_j^- \in dq | y_j^-,\S^-,\t,\T\} &\propto & q^{m_j^-}
    \e^{-\gtNtt(y_j^-)q}
    \rho(dq|y_j^-),\label{Qjn}
\end{eqnarray}
and
\begin{eqnarray}
    \hspace*{-0.3in}\eta_j(dy_j^+ |\S^+,\t,\T ) &\propto &\I(0 < y_j^+ \leq Y^\t_j)
    \kappa_{m_j^+}(\e^{-\ftNtt} \rho|y_j^+)
    \eta(dy_j^+),\label{alphaSjp}\\
    \hspace*{-0.3in}\Pr\{Q_j^+ \in dq | y_j^+,\S^+,\t,\T\} &\propto & q^{m_j^+}
    \e^{-\gtNtt(y_j^+)q}
    \rho(dq|y_j^+),\label{Qjp}
\end{eqnarray}
respectively, with existences guaranteed by~(\ref{kappa}).
\item[(iii)] Given $(\y^-, \Q^-,{\S^-}, \y^+, \Q^+,\S^+,\t,\T)$, $\mu$ has a
distribution identical to that of the random measure
$$
\mu^{\ast} = \mu_{\gtNtt} +
    \sum_{\{j^\ast|\S^-\}} Q_j^- \delta_{y_j^-}+
    \sum_{\{j^\ast|\S^+\}} Q_j^+ \delta_{y_j^+}
$$
where $\mu_{\gtNtt}$ is a completely random measure with
L\'evy measure
$
    \e^{-\gtNtt(u)x} \rho(dx|u) \eta(du).
$
\end{itemize}
}
\end{thm}

\begin{proof} When $\t$ is given, Theorem~4.1
in James~(2005) specializes and yields that the law of
$\mu|\t,\T$ can be described as the random measure
$\mu_{\gtNtt}+\sum_{i=1}^{\np}J_j \delta_{v_i}$ mixed over
by the law of $\J,\vv,\p|\t,\T$, where
$\J=(J_1,\ldots,J_{\np})$, $\vv=(v_1,\ldots,v_{\np})$
denotes the unique values of $(u_1,\ldots,u_m)$, and
$\mu_{\gtNtt}$ is a completely random measure characterized
by L\'evy measure $\e^{-\gtNtt(u)x} \rho(dx|u) \eta(du)$
with law denoted by $\CP(d\mu_{\gtNtt})$. That is, it can
be determined by the joint distribution of
$\mu_{\gtNtt},\J,\vv,\p|\t,\T$, which is proportional to
$\CP(d\mu_{\gtNtt})$ multiplies
\begin{equation}
    \prod_{i=1}^{n(\p)} {J_i}^{e_i}
    \e^{-\gtNtt(v_i) J_i} \rho(dJ_i | v_i)
    \prod_{k \in C_i}
    [\I(T_k-\t \leq v_i < 0) + \I(0 < v_i \leq T_k-\t)]
    \eta(dv_i).\label{jointp}
\end{equation}

Rewriting $\T$ as $\Z^\t$ and $\Y^\t$ as defined
in~(\ref{obs2}) and simplifying the sums of two indicators
due to~(\ref{twocases}) reveal that the $m-n$ negative
observations $\Z^\t$ can ``cluster'' only with one another
but not with any of the positive observations $\Y^\t$, or
vice versa. Hence, it is eligible to ``split'' $\p$ into
two non-overlapping partitions $\p^{-}$ and $\p^{+}$. Write
$\p = \p^- \cup \p^+$. Without loss of generality, let
$\p^{-} = \{C_1,\ldots,C_{n(\p^-)}\}$ and $\p^{+} =
\{C_{n(\p^-)+1},\ldots,C_{\np}\}$ denote the partition of
the $m-n$ negative observations $\Z^\t$ and that of the
remaining $n$ positive observations $\Y^\t$ in relation to
negative and positive unique values in $\vv$, respectively.
Hence, the law of $\J,\vv,\p|\t,\T$, proportional
to~(\ref{jointp}), becomes
\begin{eqnarray}
    &&\hspace*{-0.3in}
    \prod_{i=1}^{n(\p^-)} \left[{J_i}^{e_i}
    \e^{-\gtNtt(v_i) J_i} \rho(dJ_i | v_i)
    \I(\max_{k \in C_i}Z_k^\t \leq v_i < 0) \eta(dv_i)\right]\nonumber\\
    &&\hspace*{-0.3in}\qquad \qquad  \times
    \prod_{i=n(\p^-)+1}^{n(\p)} \left[{J_i}^{e_i}
    \e^{-\gtNtt(v_i) J_i} \rho(dJ_i | v_i)
    \I(0 < v_i \leq \min_{k \in C_i}Y_k^\t)
    \eta(dv_i)\right].
    \label{jointp2}
\end{eqnarray}
Due to its dependence on the maximal index but not the
remaining indices of each cell in both $\p^-$ and $\p^+$,
this can be represented in terms of the intrinsic
characteristics of two paths $\S^-$ and $\S^+$ of
respectively $m-n+1$ and $n+1$ coordinates via relabeling
of $\{(v_1,J_1),\ldots,(v_{n(\p^-)},J_{n(\p^-)})\}$ and
$\{(v_{n(\p^-)+1},J_{n(\p^-)+1}),\ldots,(v_{n(\p)},J_{n(\p)})\}$
respectively as $(\y^-,\Q^-)$ and $(\y^+,\Q^+)$ according
to $\p^- \in {\C}_{\S^-}$ and $\p^+ \in {\C}_{\S^+}$,
together with equalities,
\begin{eqnarray*}
\prod_{i=1}^{n(\p^-)} \I(\max_{k \in C_i} Z_k^\t \leq v_i < 0) =
\prod_{i=1}^{n(\p^-)} \I(Z_{\max_{k \in C_i} k}^\t \leq v_i < 0)
= \prod_{\{j^\ast|\S^-\}} \I(Z_j^\t \leq y_i^- < 0)\hspace*{0.3in}&&\\
\noalign{\noindent and} \prod_{i=n(\p^-)+1}^{n(\p)} \I(0 < v_i \leq
\min_{k \in C_i} Y_k^\t) = \prod_{i=n(\p^-)+1}^{n(\p)} \I(0 < v_i
\leq  Y_{\max_{k \in C_i}k}^\t) = \prod_{\{j^\ast|\S^+\}} \I(0 <
y_i^+ \leq Y_j^\t).&&
\end{eqnarray*}
That is,~(\ref{jointp}) or~(\ref{jointp2}) can be equivalently
expressed as
\begin{eqnarray}
    &&\hspace*{-0.3in}
    \prod_{\{j^{\ast}|\S^-\}} \left\{({Q_j^-})^{m_j^-}
    \e^{-\gtNtt(y_j^-) Q_j^-} \rho(dQ_j^- | y_j^-)
    \I(Z_j^\t \leq  y_j^- < 0)\eta(dy_j^-)\right\}\nonumber \\
    &&\hspace*{-0.2in}\qquad  \times
    \prod_{\{j^{\ast}|\S^+\}} \left\{({Q_j^+})^{m_j^+}
    \e^{-\gtNtt(y_j^+) Q_j^+} \rho(dQ_j^+ | y_j^+)
    \I(0 < y_j^+ \leq Y_j^\t) \eta(dy_j^+)\right\}.
    \label{jointS}
\end{eqnarray}
In other words, the law of $\mu_{\gtNtt},\J,\vv,\p|\t,\T$
only depends on $\p$ through $\S^-$ and $\S^+$. The above
equality of~(\ref{jointp}) and~(\ref{jointS}) together with
the following relation of equivalence in distribution
between the two random measures,
\begin{equation}
\label{mu}
    {\cal L}\left\{\left.\mu_{\gtNtt} +
    \sum_{i=1}^{n(\p)} J_i \delta_{v_i}\right|\t,\T\right\}
    \stackrel{d}{=} {\cal L}\left\{\left.\mu_{\gtNtt} +
    \sum_{\{j^\ast|\S^-\}} Q_j^- \delta_{y_j^-}+
    \sum_{\{j^\ast|\S^+\}} Q_j^+ \delta_{y_j^+}\right|\t,\T\right\},
\end{equation}
imply that the law of $\mu|\t,\T$ can be described as the
random measure $\mu^{\ast}$ at the right-hand side above
mixed over by the law of $\Q^-, \y^-, \S^-, \Q^+, \y^+,
\S^+|\t,\T$, which is proportional to
\begin{eqnarray}
    &&\hspace*{-0.5in}
    |{\C}_{\S^-}|\prod_{\{j^{\ast}|\S^-\}} \left\{({Q_j^-})^{m_j^-}
    \e^{-\gtNtt(y_j^-) Q_j^-} \rho(dQ_j^- | y_j^-)
    \I(Z_j^\t \leq  y_j^- < 0)
    \eta(dy_j^-)\right\}\nonumber \\
    &&\hspace*{-0.5in}  \times
    |{\C}_{\S^+}|\prod_{\{j^{\ast}|\S^+\}} \left\{({Q_j^+})^{m_j^+}
    \e^{-\gtNtt(y_j^+) Q_j^+} \rho(dQ_j^+ | y_j^+)
    \I(0 < y_j^+ \leq Y_j^\t) \eta(dy_j^+)\right\}
    \label{jointSQ}
\end{eqnarray}
and obtained by summing over all $\p^- \in {\C}_{\S^-}$ and
$\p^+ \in {\C}_{\S^+}$ in~(\ref{jointS}).
Now, the laws given by~(\ref{jointSn}-\ref{Qjp}), together
with the conditional independence relationships among them,
follow from Bayes' theorem and multiplication rule,
completing the proof.
\end{proof}

\begin{cor}\label{cor1}
{\rm The posterior mean of the BFRs in~(\ref{bathtub})
given $\t$ and $\T$ is given by, for $t\in \left[ 0,\tau
\right]$,
\begin{equation}\label{ES}
    \E[ \lambda( t|  \mu,\theta ) | \theta,\T ] =
    \sum_{\S^-} \sum_{\S^+} \,a_\lambda(t|\S^-,\S^+,\theta,\T)
     \, W(\S^-,\S^+|\t,\T)
\end{equation}
where $\sum_{{\S}}$ represents summing over all paths $\S$
of the same number of coordinates,
$$
W(\S^-,\S^+|\t,\T) = W^-(\S^-|\theta,\T)\times
W^+(\S^+|\theta,\T)
$$
is the conditional distribution of $(\S^-,\S^+)$ given $\t$
and $\T$, and
\begin{eqnarray}
  a_\lambda(t|\S^-,\S^+,\theta,\T)\hspace*{-0.2in}& &=
  \left[ \int_{t-\t}^{0}
\kappa_{1}(\e^{-\ftN} \rho|y) \eta(dy)
    +\sum_{\{j^{\ast}|\S^-\}} \lambda_{\theta,j}^-(t|
    \S^-)\right] \I(t < \theta)\hspace*{0.5in}\nonumber\\
   && \quad +\left[ \int_{0}^{t-\t} \kappa_{1}(\e^{-\ftN} \rho|y) \eta(dy)
    +\sum_{\{j^\ast|\S^+\}} \lambda_{\theta,j}^+(t|
    \S^+)\right] \I(t > \theta),\nonumber
\end{eqnarray}
wherein $\lambda_{\theta,j}^-(t | \S^-) =
\displaystyle\int_{\max(t-\t,Z_j^\t)}^{0}
\kappa_{m_j^-+1}(\e^{-\ftNtt} \rho|y)
\eta(dy)/\displaystyle\int_{Z_j^\t}^{0}
\kappa_{m_j^-}(\e^{-\ftNtt} \rho|y) \eta(dy),$ for
$j=1,\ldots,\break m-n$, and $\lambda_{\theta,j}^+(t | \S^+) =
\displaystyle\int_{0}^{\min(t-\t,Y_j^\t)}
    \kappa_{m_j^++1}(\e^{-\ftNtt} \rho|y) \eta(dy)/\displaystyle\int^{Y_j^\t}_{0}
    \kappa_{m_j^+}(\e^{-\ftNtt} \rho|y)
    \eta(dy),
$ for $j=1,\ldots,n$. }
\end{cor}

\begin{proof}
If $\u=(u_1,\ldots,u_m)$, the posterior mean of $\mu$ given
$(\u,\t,\T)$ follows from Theorem~\ref{Prop1} as
\begin{eqnarray*}
&&E[\mu^{\ast}(du)|\u,\t,\T] =
E[\mu^{\ast}(du)|\y^-,\S^-,\y^+,\S^+,\t,\T]\\
&&\qquad = \kappa_{1}(\e^{-\ftNtt} \rho|u) \eta(du) +
\sum_{\{j^{\ast}|\S^-\}} E[Q^-_j|y^-_j] \delta_{y^-_j}(du)
+ \sum_{\{j^{\ast}|\S^+\}} E[Q^+_j|y^+_j]
\delta_{y^+_j}(du),
\end{eqnarray*} where $E[Q^-_j|y^-_j] = \kappa_{m_j^-+1}(\e^{-\ftNtt}
\rho|y^-_j)/\kappa_{m_j^-}(\e^{-\ftNtt} \rho|y^-_j) $
and\\$E[Q^+_j|y^+_j] = \kappa_{m_j^++1}(\e^{-\ftNtt}
\rho|y^+_j)/\kappa_{m_j^+}(\e^{-\ftNtt} \rho|y^+_j).$
Hence, the posterior mean of $\lambda( t|  \mu,\theta )$
given $\t$ and $\T$ is
\begin{eqnarray*}
&&\hspace*{-0.3in}\sum_{\S^-}\sum_{\S^+}\left\{\int_{\R}
[\I(t-\t \leq u < 0) + \I(0 < u \leq t-\t)]
\kappa_{1}(\e^{-\ftNtt} \rho|u)
\eta(du)\right. \\
&&\qquad + \sum_{\{j^{\ast}|\S^-\}} \int_{\R} \I(t-\t \leq
y^-_j < 0) E[Q^-_j|y^-_j]
\eta^-(dy^-_j|\S^-,\t,\T)\\
&&\qquad+ \sum_{\{j^{\ast}|\S^+\}} \left.\int_{\R} \I(0 <
y^+_j \leq t-\t) E[Q^+_j|y^+_j]
\eta^+(dy^+_j|\S^+,\t,\T)\right\} W(\S^-,\S^+|\t,\T)
\end{eqnarray*}
and the result follows by comparing between $t$ and $\t$.
\end{proof}

\begin{rem}\label{mono}
{\rm When $\theta = 0$ or $\t=\infty$, Theorem~\ref{Prop1}
and Corollary~\ref{cor1} reduce to a characterization of
the posterior distribution and the posterior mean of the
class of MFRs discussed in Ho~(2006a) via one single
$\S$-path.
}
\end{rem}

With the following posterior consistency result, which is
an analogue of Theorem~4 in Dr\v{a}gichi and
Ramamoorthi~(2003) in this context, the consistency of the
above Bayes estimator of BFRs with a change point $\t$ can
be established via the same argument used in Corollary~1 of
Barron, Schervish and Wasserman~(1999). Suppose $\lambda_0$
is the true BFR defined in~(\ref{bathtub}), with a
corresponding density function $f_0$.

\begin{thm}\label{consist}
{\rm Suppose $\t$ is known and that
$\max(\displaystyle\lim_{t \rightarrow 0}
E[\lambda(t|\mu,\t)],\displaystyle\lim_{t \rightarrow
\infty} E[\lambda(t|\mu,\t)]) < \infty$ in~(\ref{bathtub}).
If $\lambda_0$ is bounded with
$\lambda_0(\t_-|\mu,\t),\lambda_0(\t_+|\mu,\t)>0$, weak
consistency holds at $f_0$.
 }
\end{thm}

\begin{proof} The proof follows from that of Theorem~4 in Dr\v{a}gichi
and Ramamoorthi~(2003) by splitting the argument based on
an increasing hazard rate on $(0,\infty)$ into two parallel
situations with respect to $\t$, as there are two
increasing hazard rates away from $\t$ of which one is
increasing from $\t$ to $\infty$ and the other one is
increasing from $\t$ to $0$.
\end{proof}

Remark~2.7 in Ho~(2006c) explains that the above
characterization of the posterior distribution and the
estimator~(\ref{ES}) for models in~(\ref{bathtub}) based on
two $\S$-paths result in significant improvements in terms
of complexity, compared with the counterparts in terms of
partitions from the general result of James~(2005). More
importantly, dividing~(\ref{jointS}), which is the joint
distribution of $(\J,\vv,\p)$ given $\t$ and $\T$,
by~(\ref{jointSQ}), the joint distribution of
$(\Q^-,\y^-,\S^-, \Q^+,\y^+,\S^+)$ given $\theta$ and $\T$,
yields the following analogue of Corollary~2.4 in
Ho~(2006c) which states that given $(\S^-,\S^+,\theta,\T)$,
$\p$ is uniformly distributed over all partitions that can
be split into $\p^-$ and $\p^+$ of which correspond to the
respective paths $\S^-$ and ${\S^+}$. Consequently, the
results in Theorem~\ref{Prop1} and Corollary~\ref{cor1},
which follow from the same argument as in Ishwaran and
James~(2003) or Ho~(2006a) to be always less variable than
their counterparts in terms of $\p$, are worthy of study
due to the posterior consistency result.

\begin{cor}\label{cor2}
{\rm Consider models in~(\ref{bathtub}). Suppose
$\S^-,\S^+|\theta,\T \sim W(\S^-,\S^+|\theta,\T)$. Then,
there exists a conditional distribution
$$
    \pi(\p|\S^-,\S^+,\theta,\T) = \frac{1}{|{\mathbb{C}}_{\S^-}||{\mathbb{C}}_{\S^+}|
    }, \qquad \p = \p^- \cup \p^+,\p^- \in
    {\C}_{\S^-},\p^+ \in
    {\C}_{\S^+},
$$
where $|\mathbb{C}_{\S^-}|$ and $|\mathbb{C}_{\S^+}|$ are
defined in~(\ref{card}). }
\end{cor}

\begin{thm} \label{combin}
{\rm Suppose the likelihood of the data $\T$ given
$(\mu,\theta)$ is proportional to~(\ref{like}). Assume that
$\mu$ is a completely random measure with L\'evy
measure~(\ref{intensity}) and the prior of $\t$ is
$\pi(d\theta)$. The posterior distribution of $\t$ is
characterized by, for any Borel set $B \in \Rp$,
\begin{equation}
  \Pr(\theta \in B|\T) = \int_B \sum_{\S^-} \sum_{\S^+}
  \pi(\S^-,\S^+,d\theta|\T)\label{posttheta},
\end{equation}
where
\begin{equation}\label{jointAll}
\pi(\S^-,\S^+,d\theta|\T) \propto {\cal
L}_{\mu}(g_{N,\t}|\rho, \eta)
\phi^-_{\theta}({\S^-},\T)\,\phi^+_{\theta}({\S^+},\T)
\,\pi(d\theta)
\end{equation}
defines a joint distribution of $(\S^-,\S^+,\t)$ given
$\T$, with a normalizational constant\break $\int_{\Rp}
{\cal L}_{\mu}(g_{N,\vartheta}|\rho, \eta) \sum_{\S^+}
\sum_{\S^-}\phi^-_{\vartheta}({\S^-},\T)\,
\phi^+_{\vartheta}({\S^+},\T)\,\pi(d\vartheta)$ and ${\cal
L}_{\mu}(\cdot|\rho, \eta)$, $\phi^-_\theta({\S^-},\T)$ and
$\phi^+_\theta({\S^+},\T)$ defined
in~(\ref{Lap}),~(\ref{jointSn}) and~(\ref{jointSp}),
respectively.}
\end{thm}

\begin{proof}
Applying Proposition~2.1 in James~(2005) and following the
same argument as in proving Theorem~\ref{Prop1} yield a
joint distribution of $(\J,\vv,\p,\t)$ given $\T$, which is
proportional to the expression~(\ref{jointp}) multiplies
${\cal L}_{\mu}(g_{N,\t}|\rho, \eta)\pi(d\t)$. Integrating
$(\J,\vv)$, which is equivalent to integrating $(\Q^-,
\y^-, \Q^+, \y^+)$ in~(\ref{jointS}), gives a joint
distribution of $(\S^-,\S^+,\t)$ given $\T$ as
in~(\ref{jointAll}). Result follows from further
marginalization of $(\S^-,\S^+)$.
\end{proof}

When $\theta$ is not known, posterior analysis of models
in~(\ref{bathtub}) follows from~(\ref{double}) with
$\CP(d\theta|\T)$ defined above. For instance, the
posterior mean of hazard rates in~(\ref{bathtub}) given
$\T$ is given by
\begin{equation}\label{Ef}
    \E[ \lambda( t| \mu,\theta ) |\T ] = \int_{\Rp} \sum_{\S^-} \sum_{\S^+}
    a_\lambda(t|\S^-,\S^+,\theta,\T) \pi(\S^-,\S^+,d\theta|\T),
\end{equation}
where $a_\lambda(t|\S^-,\S^+,\theta,\T)$ is defined in
Corollary~\ref{cor1}.

\section{Monte Carlo procedures\label{sec:MC}}
This section introduces Monte Carlo procedures for
evaluating/approximating posterior quantities of models
in~(\ref{bathtub}), like~(\ref{ES}),~(\ref{posttheta})
and~(\ref{Ef}), which are expressible as finite sums over
two $\S$-paths, based on sampling the triplets $( \S^-,
\S^+,\theta)$ in light of the data $\T$. For brevity,
conditioning statements on the data $\T$ will be suppressed
throughout in this section as all sampling procedures are
designed with respect to distributions conditioning on
$\T$. Firstly, when $\t$ is given, both iterative and
non-iterative procedures for sampling the paths $( \S^-,
\S^+)$ will be discussed. Then, a sequential importance
sampling~(SIS) scheme for drawing the triplets from the
posterior distribution $\pi(\S^-,\S^+,d\t|\T)$
in~(\ref{jointAll}) is proposed. Conditional independence
between $\S^-$ and $\S^+$ given $\t$ and $\T$ stated in
statement~(i) of Theorem~\ref{Prop1}, the nice structure of
the posterior distribution for models in~(\ref{bathtub}),
plays a crucial role in constructing all the algorithms
that follow.

\subsection{When $\t$ is known}
\subsubsection{A Gibbs sampler}\label{sec:MCMC}

Define a generalization of the accelerated path~(AP)
sampler introduced in Ho~(2002)~(see also Ho~(2006a,b)),
which is an efficient MCMC algorithm for sampling one
single $\S$-path at a time in the context of Bayes
estimation of monotone hazard rates and monotone densities,
as follows.

\begin{algorithm}[The AP sampler]\label{AP}
{\rm A Markov chain of $\S$-paths of $n+1$ coordinates with
a unique stationary distribution,
\begin{equation}\label{piS} 
\pi(\S)
\propto \phi(\S) 
= |\C_{\S}| \prod_{\{j^*|\S\}} \psi^{(m_j)}(X_j),
\end{equation}
where $\psi^{(m_j)}(X_j)$ is a finite real-valued function depending
on $m_j$ and $X_j$ only, and $X_1,\ldots,X_n$ is a
decreasing/increasing sequence in $\R$, can be defined by a
transition cycle of $n-1$ steps:

\begin{itemize}
\item[(I)] At step $r$, suppose ${\S}^{\ast} =
(0,S_1,\ldots,S_{r-1}, c, \ldots, c, S_q, \ldots, S_{n-1}, n)$,
where $S_{r-1} \leq c \leq \mbox{min}(r,S_{q}-1)$ and $q>r$ denotes
the next location at which $m_q = S_q - S_{q-1}
> 0$. The chain moves from ${\S}^{\ast}$ to $\S^{\ast\ast}_{r,q,k} =
(0,S_1,\ldots,S_{r-1},k, \ldots, k, S_q, \ldots, S_{n-1},n)$ with
conditional probability proportional to $
\phi(\S^{\ast\ast}_{r,q,k})$ for $k = S_{r-1}, S_{r-1}+1, S_{r-1}+2,
\ldots, \mbox{min}(r,S_q - 1)$.
\item[(II)] Repeat step~(I) for $r=1, 2, \ldots, n-1$ to complete a
cycle.
\end{itemize}
}
\end{algorithm}

Starting with an arbitrary path ${\S}_{(0)}$, and repeating
$M$ cycles according to the above scheme, give a Markov
chain ${\S}_{(0)}, {\S}_{(1)}, \ldots, {\S}_{(M)}$ with a
unique stationary distribution $\pi({\S})$. 
We remark that the sequence of determination of coordinates
$S_i$ in the AP sampler does not have much effect on its
effectiveness or efficiency.

As a consequence of conditional independence between $\S^-$
and $\S^+$ given $\t$ and $\T$, an iterative scheme, dubbed
as \textit{accelerated paths~(APs) sampler}, for sampling a
pair of $(\S^-,\S^+)$ from the posterior distribution
$W(\S^-,\S^+|\theta,\T) = W^-(\S^-|\t,\T) \times
W^+(\S^+|\t,\T)$ in Corollary~\ref{cor1} can be defined
naturally by two independent implementations of the AP
sampler, or, by cycling through the following two steps in
a cycle:
\begin{enumerate}
    \item[(M1)] Determine $\S^-$ by applying Algorithm~\ref{AP}
    with $n$, $\phi(\S)$, $X_1,\ldots,X_n$ and
    $\psi^{(m_j)}(X_j)$ replaced by $m-n$, $\phi^-_\t(\S^-,\T)$,
    $Z^\t_1,\ldots,Z^\t_{m-n}$ and $\int_{Z^\t_j}^{0}
    \kappa_{m_j^-}(\e^{-\ftNtt} \rho|y) \eta(dy)$, respectively.
    \item[(M2)] Determine $\S^+$ by applying
    Algorithm~\ref{AP} with $\phi(\S)$, $X_1,\ldots,X_n$ and
    $\psi^{(m_j)}(X_j)$ replaced by $\phi^+_\t(\S^+,\T)$,
    $Y_1^\t,\ldots,Y_{n}^\t$ and $\int^{Y^\t_j}_{0}
    \kappa_{m_j^+}(\e^{-\ftNtt} \rho|y) \eta(dy)$, respectively.
\end{enumerate}

A Markov chain $({\S}_{(0)}^-,\S_{(0)}^+),
({\S}_{(1)}^-,{\S}_{(1)}^+), \ldots,
({\S}_{(M)}^-,{\S}_{(M)}^+)$ with a unique stationary
distribution $W(\S^-,\S^+|\t,\T)$ can be obtained by
starting with an arbitrary pair of paths ${\S}_{(0)}^-$ and
$\S^+_{(0)}$, and repeating $M$ cycles of steps~(M1)
and~(M2). Then, expectations of any functional
$h(\S^-,\S^+)$ with respect to the probability distribution
$W(\S^-,\S^+|\theta,\T)$ can be approximated by the ergodic
average~[Meyn and Tweedie~(1993)]
$$
\nu_{h,\t}^{M} = \frac{1}{M} \sum_{i=1}^M h(\S_{(i)}^-,\S_{(i)}^+).
$$
For instance, the posterior mean $\E[ \lambda( t| \mu,\theta ) |
\theta,\T ]$ in~(\ref{ES}) can be approximated by
\begin{equation}\label{MCMC1hr}
\nu_{a_\lambda,\t}^{M}(t) = \frac{1}{M} \sum_{i=1}^M
a_\lambda(t|\S_{(i)}^-,\S_{(i)}^+,\theta,\T).
\end{equation}

\subsubsection{A sequential importance sampling method}\label{sec:SIS}
Due to the same reason as for constructing the APs sampler,
we propose an SIS~[Kong, Liu and Wong~(1994) and Liu and
Chen~(1998)] method for sampling the two paths from
$W(\S^-,\S^+|\theta,\T)$ which is designed as two
independent implementations of an SIS scheme for sampling
one path at a time, called the \textit{sequential
importance path~(SIP) sampler} introduced in Ho~(2006c).
The SIP sampler is an SIS scheme that allows us to draw an
$\S$-path of $n+1$ coordinates according to a probability
distribution $\pi(\S) \propto \phi(\S)$ defined
by~(\ref{piS}). Let $I_{0}=0$ and $I_{n}=n$.

\begin{algorithm}[The SIP sampler in Ho~(2006c)]\label{SIP} {\rm Based on a random
permutation $\Xi_{n-1}= \{I_{1},\ldots,I_{n-1}\}$ of the
integers $\{1,2,\ldots,n-1\}$, an SIS method for sampling
an $\S$-path of $n+1$ coordinates from $\pi(\S)$ given
in~(\ref{piS}) consists of recursive applications of the
following SIS steps for $r=1,\ldots,n-1$:
\begin{enumerate}
    \item[A.] Given $D_{r-1} \equiv
    \{I_0\} \cup \{I_1,\ldots,I_r\} \cup \{I_n\}$, which is
    the collection of all indices $i$ whereby $S_i$ has been determined up to step $r-1$, let
    $p=\max\{I_{j} \in D_{r-1}:I_{j}<I_r\}$ and
$q=\min\{I_{j}\in D_{r-1}:I_{j}>I_r\}$. Determine $S_{I_r} = k$, for
$k = S_p, S_p+1, \ldots, \min(I_r,S_q)$, according to a probability
distribution
    $$
    \sigma_{r}(k|\{S_h:h \in D_{r-1}\})
    \propto \phi(\S_{I_r,k}^{\ast}),
    $$
    where $\S_{I_r,k}^{\ast} = (0,S^{\ast}_1,\ldots,S^{\ast}_{I_r-1},S^{\ast}_{I_r},
    S^{\ast}_{I_r+1},\ldots,S^{\ast}_{n-1},n)$ is a path of $n+1$ coordinates
    such that $S_{I_r}^{\ast} = k$ and for $i=1,\ldots,I_{r}-1,I_{r}+1,\ldots,n-1$,
    $S_i^{\ast} = S_{I_h}$ if $i = I_h \in D_{r-1}$; otherwise, $S_i^{\ast}
    = S_{i-1}^{\ast}$.
    \item[B.] Compute $\sigma_{r}(k|\{S_h:h \in D_{r-1}\})$, which equals
    $\phi(\S^\ast_{I_r,k})$
    multiplied by the appropriate constant of proportionality, for the chosen
    value $k$ of $S_{I_r}$.
\end{enumerate}
}
\end{algorithm}

After step $n-1$, a random path $\S =
(0,S_1,S_2,\ldots,S_{n-1},n)$ distributed as
\begin{equation}\label{sigma}
\sigma_{n-1}(\S) = \prod_{r=1}^{n-1} \sigma_{r}(S_{I_r}|\{S_h:h \in
D_{r-1}\})
\end{equation}
can be obtained. The importance sampling weight of this
realized path $\S$ is given by $\upsilon_{n-1}(\S) =
\phi(\S) / \sigma_{n-1}(\S)$. Or, $\S$ is said to be
\textit{properly weighted} by a weighting function
$\upsilon_{n-1}(\S)$ with respect to the distribution
$\pi(\S)$ in~(\ref{piS})~[Liu and Chen~(1998)].

\begin{algorithm}[Sequential
importance paths~(SIPs) sampler]\label{SIP2} {\rm For a fixed value
of $\t$, an SIS method for sampling a random pair of $(\S^-,\S^+)$
from the posterior distribution $W(\S^-,\S^+|\theta,\T)$ consists of
the following three steps:
\begin{itemize}
    \item[(S1)] Obtain $\Z^\t$ and $\Y^\t$ based on $\t$ according
    to~(\ref{obs2}). Get random permutations $\Xi_{m-n-1}$
    and $\Xi_{n-1}$ of the integers $\{1,\ldots,m-n-1\}$ and
    $\{1,\ldots,n-1\}$, respectively.
    \item[(S2)] Determine $\S^-$ of $m-n+1$ coordinates by
    applying Algorithm~\ref{SIP} based on $\Xi_{m-n-1}$
    with $n$, $\phi(\S)$, $X_1,\ldots,X_n$ and
    $\psi^{(m_j)}(X_j)$ replaced by $m-n$, $\phi^-_\t(\S^-,\T)$,
    $Z^\t_1,\ldots,Z^\t_{m-n}$ and $\int_{Z^\t_j}^{0}
    \kappa_{m_j^-}(\e^{-\ftNtt} \rho|y)
    \eta(dy)$, respectively. Obtain $\sigma_{m-n-1}(\S^-|\theta)$
    according to~(\ref{sigma}).
    \item[(S3)] Determine $\S^+$ of $n+1$ coordinates by applying
    Algorithm~\ref{SIP} based on $\Xi_{n-1}$
    with $\phi(\S)$, $X_1,\ldots,X_n$ and
    $\psi^{(m_j)}(X_j)$ replaced by $\phi^+_\t(\S^+,\T)$,
    $Y_1^\t,\ldots,Y_{n}^\t$ and\break $\int^{Y^\t_j}_{0}
    \kappa_{m_j^+}(\e^{-\ftNtt} \rho|y)
    \eta(dy)$, respectively. Obtain $\sigma_{n-1}(\S^+|\theta)$
    according to~(\ref{sigma}).
\end{itemize}
}
\end{algorithm}

The pair $(\S^-,\S^+)$ is said to be \textit{properly weighted} by a
weighting function
$$
\omega_{m-2,\t}(\S^-,\S^+) = \frac{\phi_\t^-(\S^-,\T)
\phi_\t^+(\S^+,\T)}
{\sigma_{m-n-1}(\S^-|\theta)\sigma_{n-1}(\S^+|\theta)},
$$
wherein $m-2$ in the subscript representing the total
number of SIS steps, with respect to
$W(\S^-,\S^+|\theta,\T)$. Note that steps~(S2) and~(S3)
above are interchangeable as the two paths are
conditionally independent given $\t$. Replicating the above
algorithm $M$ times gives $M$ iid pairs of draws,
$(\S_{(1)}^-,\S_{(1)}^+),\ldots,(\S_{(M)}^-,\S_{(M)}^+)$,
with respective importance sampling weights,\break
$\omega_{m-2,\t}(\S_{(1)}^-,\S_{(1)}^+), \ldots,
\omega_{m-2,\t}(\S_{(M)}^-,\S_{(M)}^+)$. Then, expectations
of any functional $h(\S^-,\S^+)$ with respect to the
probability distribution $W(\S^-,\S^+|\theta,\T)$ 
can be approximated by
$$
\eta_{h,\t}^{M} = \frac{\sum_{i=1}^M h(\S_{(i)}^-,\S_{(i)}^+)\,
\omega_{m-2,\t}(\S_{(i)}^-,\S_{(i)}^+)}{\sum_{i=1}^M
\omega_{m-2,\t}(\S_{(i)}^-,\S_{(i)}^+)}.
$$
For example, the posterior mean $\E[ \lambda( t| \mu,\theta ) |
\theta,\T ]$ in~(\ref{ES}) can be approximated by
\begin{equation}\label{SIS1hr}
\eta_{a_\lambda,\t}^{M}(t) = \frac{\sum_{i=1}^M
a_\lambda(t|\S_{(i)}^-,\S_{(i)}^+,\theta,\T)\,
\omega_{m-2,\t}(\S_{(i)}^-,\S_{(i)}^+)}{\sum_{i=1}^M
\omega_{m-2,\t}(\S_{(i)}^-,\S_{(i)}^+)}.
\end{equation}

\subsection{When $\t$ is unknown -- SIPs$(\t)$ sampler}
When $\t \in \Rp$ is unknown, we can design an SIS scheme,
dubbed as \textit{SIPs$(\t)$ sampler}, which is basically
as a slight extension of the SIPs
sampler~(Algorithm~\ref{SIP2}), for sampling the triplets
from $\pi(\S^-,\S^+,d\theta|\T)$ in~(\ref{jointAll});
inserting the following step,
\begin{itemize}
    \item[(S0)] Sample $\theta$ according to a density
    $\rho(\theta) >0$, $\theta \in \R$,
\end{itemize}
before implementing the three steps~(S1--S3) in
Algorithm~\ref{SIP2} gives a random sample of
$(\S^-,\S^+,\theta)$, which is properly weighted by a
weighting function
$$
\omega_{m-1}(\S^-,\S^+,\theta) = \frac{{\cal L}_{\mu}(g_{N,\t}|\rho,
\eta) \phi^-_\theta({\S^-},\T) \,\phi^+_\theta({\S^+},\T)
\,\pi(\theta)} {\sigma_{m-n-1}(\S^-|\theta)\,
\sigma_{n-1}(\S^+|\theta) \, \rho(\theta)}
$$
with respect to $\pi(\S^-,\S^+,d\theta|\T)$ if
$\pi(d\theta) = \pi(\theta)d\theta$. Note that the total
number of positive observations $n$ is no longer a constant
as it is in Algorithm~\ref{SIP2}; $n$, depending on $\t$,
is fixed in step~(S1) only after each determination of $\t$
in step~(S0). Suppose we implement the SIPs$(\t)$ sampler
independently for $M$ times to get $M$ iid draws of the
triplets, $(\S_{(1)}^-,\S_{(1)}^+,\t_{(1)}),\ldots,\break
(\S_{(M)}^-,\S_{(M)}^+,\t_{(M)})$, with respective
importance sampling weights,
$\omega_{m-1}(\S_{(1)}^-,\S_{(1)}^+,\t_{(1)}), \ldots,
\break\omega_{m-1}(\S_{(M)}^-,\S_{(M)}^+,\t_{(M)})$. For
any function $h(\S^-,\S^+,\t)$,
$$
\E[h(\S^-,\S^+,\t)|\T] \equiv \int_{\Rp} \sum_{\S^+} \sum_{\S^-}
h(\S^-,\S^+,\t) \pi(\S^-,\S^+,d\t|\T) \approx \eta_{h}^{M}
$$
where
$$
    \eta_{h}^{M} =\frac{\sum_{i=1}^M
h(\S_{(i)}^-,\S_{(i)}^+,\t_{(i)})\,
\omega_{m-1}(\S_{(i)}^-,\S_{(i)}^+,\t_{(i)})}{\sum_{i=1}^M
\omega_{m-1}(\S_{(i)}^-,\S_{(i)}^+,\t_{(i)})}.
$$
Hence, in Theorem~\ref{combin}, the posterior
probability~(\ref{posttheta}) can be approximated by
setting $h(\S^-,\S^+,\t) = \I(\t \in B)$, that is,
\begin{equation}\label{SIS2theta}
\Pr(\t \in B|\T) = \E[\I(\t \in B)|\T]\approx  \frac{\sum_{i=1}^M
\I(\t_{(i)} \in B)\,
\omega_{m-1}(\S_{(i)}^-,\S_{(i)}^+,\t_{(i)})}{\sum_{i=1}^M
\omega_{m-1}(\S_{(i)}^-,\S_{(i)}^+,\t_{(i)})}.
\end{equation}
Similarly, regarding the Bayes estimate of the BFRs
in~(\ref{bathtub}) given by~(\ref{Ef}), we have
\begin{equation}\label{SIS2hr}
\E[ \lambda( t| \mu,\theta ) | \T ] \approx
\eta_{a_\lambda}^{M}(t) = \frac{\sum_{i=1}^M
a_\lambda(t|\S_{(i)}^-,\S_{(i)}^+,\theta_{(i)},\T)\,
\omega_{m-1}(\S_{(i)}^-,\S_{(i)}^+,\t_{(i)})}{\sum_{i=1}^M
\omega_{m-1}(\S_{(i)}^-,\S_{(i)}^+,\t_{(i)})}.
\end{equation}

\section{Numerical Results}\label{sec:sim}

This section illustrates the methodology with numerical
examples. For purpose of illustration, $\mu$ is selected to
be a gamma process with shape measure as a uniform density
on $[-2\tau,2\tau]$, that is, a completely random measure
with L\'{e}vy measure
$$
    \rho(dx|u)\eta(du) =
     x^{-1}
    \e^{-x} \I(x > 0) \,dx \times  \frac{1}{4\tau}\I(-2\tau < u < 2\tau)
    \,du,
$$
as it results in closed and easily manageable expressions
for most quantities that appear so far. The prior
$\pi(d\theta)$ is chosen to be uniformly distributed on a
reasonably large interval on $\Rp$ to ``deflate'' the prior
belief.
Simulated data are generated from two bathtub-shaped life
distributions to test the methodology. The life
distributions correspond to BFRs given by
\begin{equation}
    \lambda_1( t) =\left\{
\begin{array}
[c]{lll}%
1, &  & 0<t\leq 0.5,\\
e^{-1}, &  & 0.5<t\leq 3,\\
e^{-2/3}, &  & t>3.
\end{array}
\right.  \label{hazard6}
\end{equation}
and
\begin{equation}
    \lambda_2( t) =\left\{
\begin{array}
[c]{lll}%
e^{-2.5 t}, &  & 0<t\leq 1,\\
e^{-2.5}, &  & 1<t\leq 5,\\
e^{-6+0.7t}, &  & t>5,
\end{array}
\right.  \label{hazard4}
\end{equation}
respectively. The censoring rates in the data sets governed
by hazard rates~(\ref{hazard6}) and~(\ref{hazard4}) are
about 15\% and 20\% by setting termination times $\tau = 4$
and $\tau = 8$, respectively. Last but not least, Monte
Carlo size $M = 10,000$ is chosen for implementations of
the proposed SIS methods in all results that follow.



\begin{figure}[ht]
  \centering\includegraphics[width=5in]{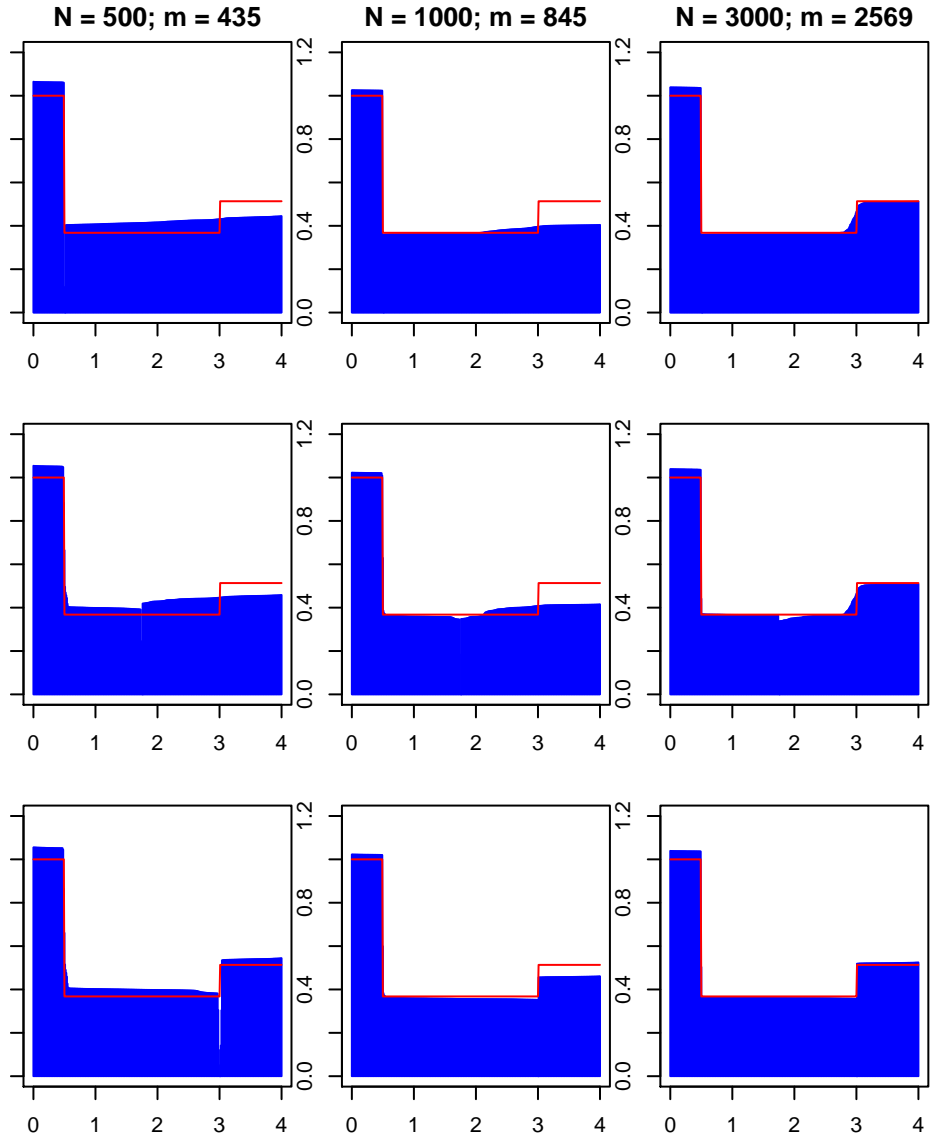}\\
\renewcommand{\baselinestretch}{1}
  \parbox{5in}
  {\caption{\quad The true bathtub-shaped hazard rate
  $\lambda_1(t)$~(solid line) given by~(\ref{hazard6}) and the Bayes estimates
  produced by the APs sampler based on total number of observations, $N =
  500$~(left column), $1,000$~(middle column) and $3,000$~(right column),
  with $\theta = 0.5, 1.75\mbox{ and } 3$~(from top row to bottom row).}
  \label{fig:AP6}}
\end{figure}

\begin{figure}[ht]
\centering\includegraphics[width=5in]{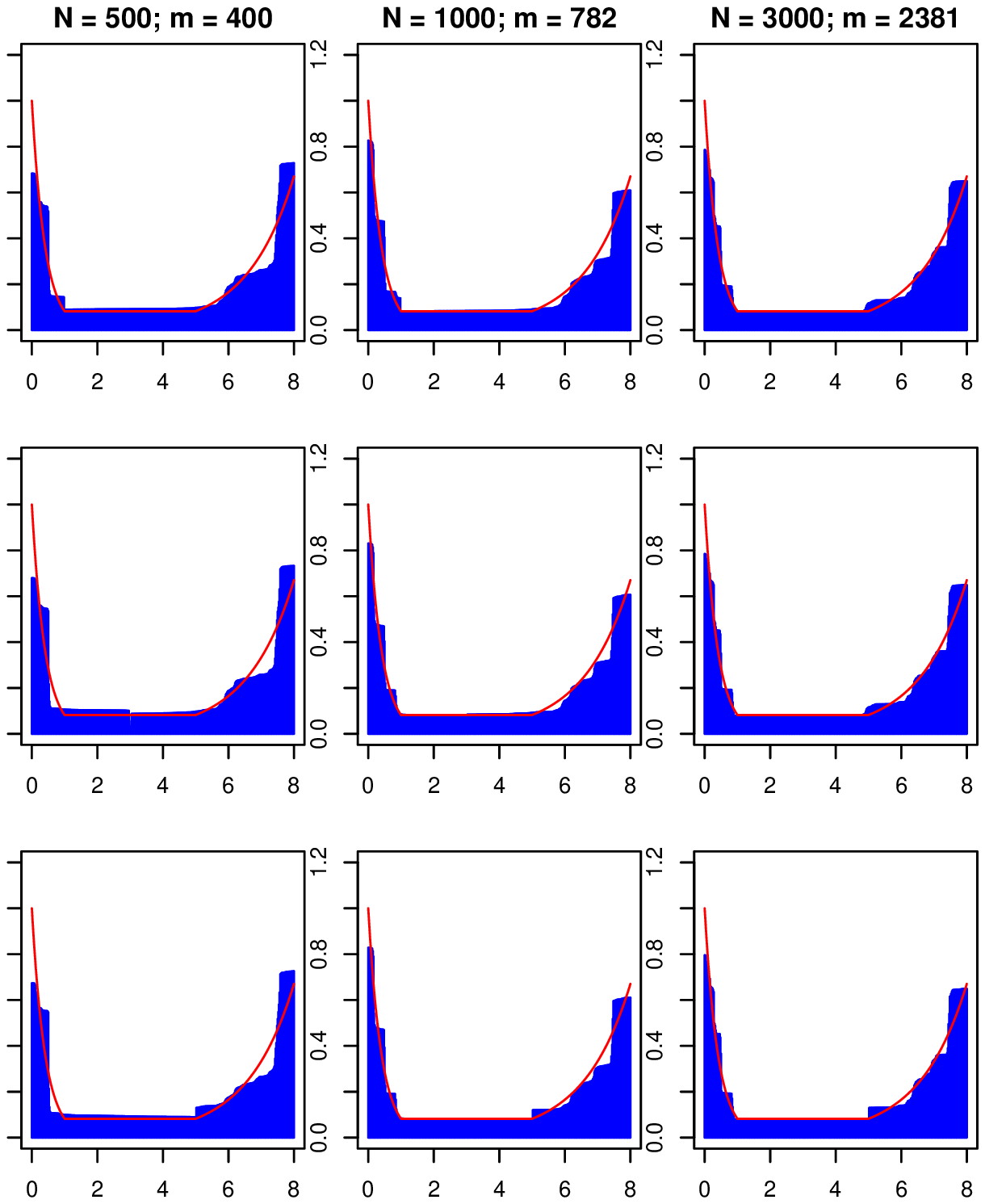}\\
\renewcommand{\baselinestretch}{1}
  \parbox{5in}
  {\caption{\quad The true bathtub-shaped hazard rate
  $\lambda_2(t)$~(solid line) given by~(\ref{hazard4}) and the Bayes estimates
  produced by the APs sampler based on total number of observations, $N =
  500$~(left column), $1,000$~(middle column) and $3,000$~(right column),
  with $\theta = 1, 3\mbox{ and } 5$~(from top row to bottom row).}
  \label{fig:AP4}}
\end{figure}

\begin{figure}[th]
  \centering\includegraphics[width=5in]{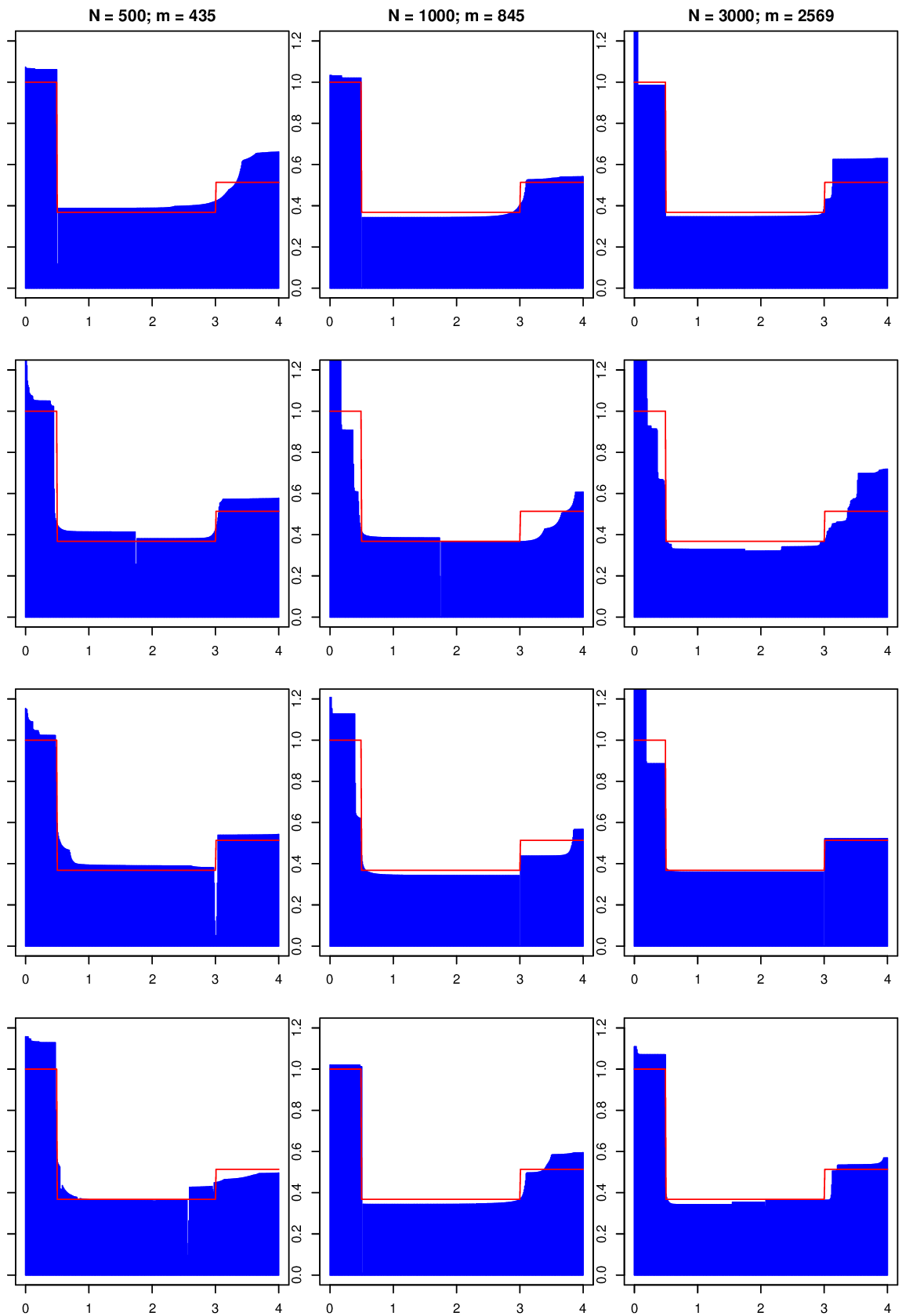}\\
\renewcommand{\baselinestretch}{1}
  \parbox{5in}
  {\caption{\quad The true bathtub-shaped hazard rate
  $\lambda_1(t)$~(solid line) given by~(\ref{hazard6}) and the Bayes estimates
  produced by the SIS methods based on total number of observations, $N =
  500$~(left column), $1000$~(middle column) and $3000$~(right column),
  wherein estimates in the first three rows from top to bottom are obtained by the SIPs
  sampler~(Algorithm~\ref{SIP2}) with $\theta = 0.5, 1.75\mbox{ and } 3$, respectively, and
  those in the last row are obtained by the SIPs$(\t)$ sampler with an unknown $\t$.}
  \label{fig:SIPS6}}
\end{figure}

\begin{figure}[ht]
\centering\includegraphics[width=5in]{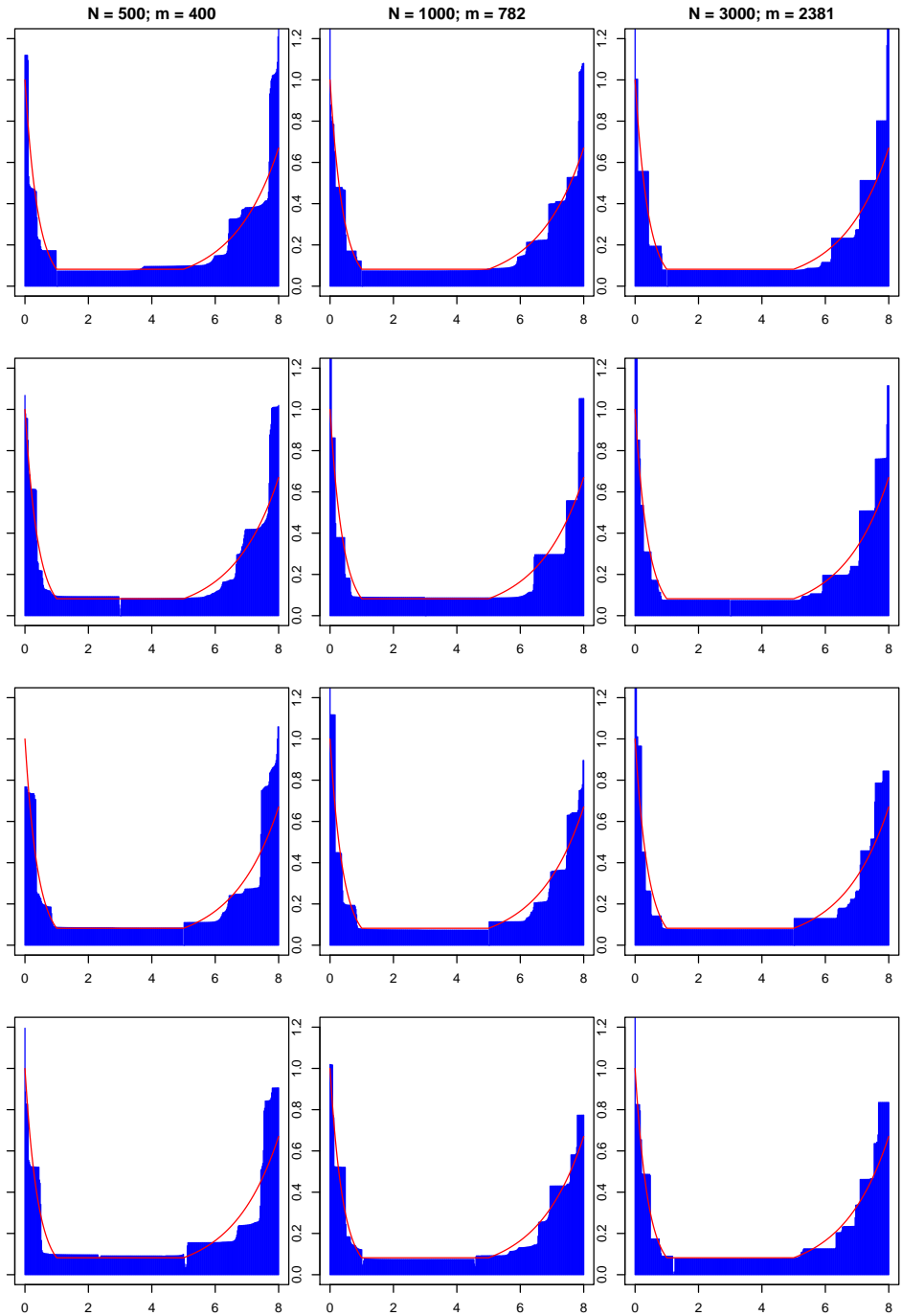}\\
\renewcommand{\baselinestretch}{1}
  \parbox{5in}
  {\caption{\quad The true bathtub-shaped hazard rate
  $\lambda_2(t)$~(solid line) given by~(\ref{hazard4}) and the Bayes estimates
  produced by the SIS methods based on total number of observations, $N =
  500$~(left column), $1000$~(middle column) and $3000$~(right column),
  wherein estimates in the first three rows from top to bottom are obtained by the SIPs
  sampler~(Algorithm~\ref{SIP2}) with $\theta = 1, 3\mbox{ and } 5$, respectively, and
  those in the last row are obtained by the SIPs$(\t)$ sampler with an unknown $\t$.}
  \label{fig:SIPS4}}
\end{figure}

Our attention is to first investigate whether the iterative
scheme and the SIS method work well when $\t$ is fixed.
The APs sampler discussed in Section~\ref{sec:MCMC} and the
SIPs sampler~(Algorithm~\ref{SIP2}) are implemented based
on a fixed value of $\t$, wherein the APs sampler is
initialized by paths ${\S}^-_{(0)}$ and ${\S}^+_{(0)}$ with
coordinates $S_i^- = S_i^+ = i$, for all $i$, to produce
totally $M = 1,000$ pairs of paths in the sense that
samples are taken once every 5 cycles after a ``burn-in''
period of $5,000$ cycles. As there is a long interval in
which the test BFRs~(\ref{hazard6}) and~(\ref{hazard4})
attain their minimum value, both the algorithms are
implemented with three different values of $\t$ in order to
see whether there is any significant effect of different
choices of $\t$ on the performance. For fitting
$\lambda_1(t)$, $\t$ is fixed at 0.5, 1.75 and 3, whereas
for fitting $\lambda_2(t)$, 1, 3 and 5 are selected. In
particular, the convergence property of the approximated
hazard rate estimates as the total number of observations
$N$ increases is studied. Figures~\ref{fig:AP6}
and~\ref{fig:AP4} depict ergodic averages~(\ref{MCMC1hr})
produced by the APs sampler with the aforementioned
different values of $\t$ based on nested samples of sizes
$N=500$, $1,000$ and $3,000$ from the life distribution
governed by BFRs~(\ref{hazard6}) and~(\ref{hazard4}),
respectively. Corresponding weighted average
estimates~(\ref{SIS1hr}) produced by the non-iterative SIPs
sampler for approximating~(\ref{ES}) are graphed in the
first three rows of Figures~\ref{fig:SIPS6}
and~\ref{fig:SIPS4}.

To investigate the performance of the SIPs$(\t)$ sampler
when $\t$ is not known, we set $\rho(\theta)$ to be uniform
on an interval which includes all the complete
observations. Independent random samples of $(\S^-, \S^+,
\t)$ of size $M=10,000$ are resulted from implementing the
sampler based on the same sets of nested samples of sizes
$N=500$, $1,000$ and $3,000$ according to the two hazard
rates $\lambda_1(t)$ and $\lambda_2(t)$. For the sake of a
better comparison between results by the SIPs$(\t)$ sampler
based on an unknown $\t$ and those by the SIPs sampler with
a fixed $\t$, the resulting Bayes estimates of the
BFRs~(\ref{hazard6}) and~(\ref{hazard4}), given by the
weighted average~(\ref{SIS2hr}), are presented in the last
rows of Figures~\ref{fig:SIPS6} and~\ref{fig:SIPS4},
respectively.

In summary, the graphs echo the fact that approximations
for Bayes estimates of the BFRs in~(\ref{bathtub}) by all
the proposed algorithms tend to the ``true'' hazard rates,
$\lambda_1(t)$ and $\lambda_2(t)$, as sample size
increases. We remark that some other simulations we have
carried out applying the APs and the SIPs samplers based on
fixed values of $\t$ other than those stated above reveal
that there is not much difference between simulation
results based on different values of $\t$.

\clearpage

\section{A Test of an MFR Versus an BFR}
\label{sec:testing}

Early references devoted to testing for a constant hazard
rate versus an MFR include Proschan and Pyke~(1967), Bickel
and Doksum~(1969) and Gail and Gastwirth~(1978a,b), among
others. Without relying on exponentiality assumption,
Gijbels and Heckman~(2004) develop a testing procedure via
normalized spacings for testing an MFR against alternatives
of some local departures. For testing an MFR versus other
general alternatives, Hall and Van Keilegom~(2005) propose
a calibration method related to the ``increasing
bandwidth'' approach suggested by Silverman~(1981) in the
case of density estimation. Testing procedures involving
BFRs can be found in, for example, Aarset~(1985), who
discussed the test statistic proposed by Bergman~(1979) for
testing a constant hazard rate against an BFR, and
Vaurio~(1999), who proposed a few test statistics for
testing between an MFR and other non-monotone alternatives
including BFRs.

A Bayesian test of monotone versus bathtub-shaped hazard
rates can be readily defined in terms of $\t$ based on the
models in~(\ref{bathtub}) with $\mu$ being a nuisance
parameter as follows: Suppose we are interested in testing
whether a set of observations $\T$, defined similarly in
Section~\ref{sec:model}, is generated according to a
non-decreasing hazard rate or an BFR. Based
on~(\ref{bathtub}), it is equivalent to choose between two
hypotheses $H_0:\t = 0$ and $H_1:\t \in (0,\infty)$ as when
$\t = 0$, models in~(\ref{bathtub}) correspond to a class
of non-decreasing hazard rates; otherwise, they give a
class of BFRs with a change point $\t>0$. In particular,
the likelihood of the data given $(\mu,\t)$ under $H_1$ is
given by~(\ref{like}) when $\t \ne 0$ or $\infty$,
while the likelihood of the data given $\mu$ under $H_0$
follows from~(\ref{like}) with $\t = 0$ as
\begin{equation}
\label{likeH0}
    \e^{-\mugtNt0}
    \prod_{i=1}^{m} 
    \int
    \I(0 < u_i \leq T_i) \mu(du_i).
\end{equation}
Let $\pi_0$ denote the prior probability of $H_0$, and then
$1-\pi_0$ denotes the prior probability of $H_1$;
furthermore, suppose the mass on $H_1$ is spread out
according to a distribution $\pi(d\t)$. Suppose we assume
that $\mu$'s under $H_0$ and $H_1$ are two independent, but
not necessarily identical, completely random measures
characterized by~(\ref{Lap}).

\begin{cor} 
{\rm Suppose $\mu$ is a completely random measure
characterized by~(\ref{Lap}). It follows from
Theorem~\ref{combin} that the likelihood of the data $\T$
given $\t$ is proportional to
\begin{equation}
m_\t(\T) = {\cal L}_{\mu}(g_{N,\t}|\rho, \eta)\times
\sum_{\S^-}\phi^-_{\t}({\S^-},\T)\times
\sum_{\S^+}\phi^+_{\t}({\S^+},\T).
\end{equation}
}
\end{cor}
Hence, the marginal density of $\T$ is given by
\begin{equation}\label{marginal}
m(\T) = \pi_0 \times {\cal L}_{\mu}(g_{N,0}|\rho, \eta)
\sum_{\S^+} \phi^+_{0}({\S^+},\T)+(1-\pi_0) \times
\int_{\Rp} m_\t(\T) \pi(d\t).
\end{equation}
It implies that the posterior probability of $H_0$ is given
by
$$
P(H_0|\T) = \frac{\pi_0\times {\cal L}_{\mu}(g_{N,0}|\rho,
\eta) \sum_{\S^+} \phi^+_{0}({\S^+},\T)}{m(\T)},
$$
and that of $H_1$ is equal to $1-P(H_0|\T)$. Also of
interest is the posterior odds of $H_0$ to $H_1$, which is
given by
$$
\frac{\pi_0}{1-\pi_0} \times \frac{{\cal
L}_{\mu}(g_{N,0}|\rho, \eta) \sum_{\S^+}
\phi^+_{0}({\S^+},\T)}{\int_{\Rp} m_\t(\T) \pi(d\t)},
$$
wherein $\pi_0/(1-\pi_0)$ is the prior odds and the latter
ratio is the Bayes factor for $H_0$ versus $H_1$~(see Kass
and Raftery~(1995) for a review of Bayes factors).

Regarding implementation of the above Bayesian test,
Algorithm~\ref{SIP} and the SIP$(\t)$ sampler can be
applied to approximate the marginal density of $\T$,
$m(\T)$, in~(\ref{marginal}), and also the posterior
probabilities of $H_0$ and $H_1$. On one hand, the sum
$\sum_{\S^+} \phi_0^+(\S^+,\T)$ is approximated by
$$
\frac{1}{M}\sum_{i=1}^M \sigma_{m-1}(\S_{(i)}),
$$
if ${\S}_{(0)}, {\S}_{(1)}, \ldots, {\S}_{(M)}$ are
independent samples obtained via implementing
Algorithm~\ref{SIP} with $\phi(\S) = \phi_0^+(\S,\T)$
in~(\ref{piS}) and $\sigma_{m-1}(\S_{(i)})$ defined
in~(\ref{sigma}). On the other hand, the integral
$\int_{\Rp} m_\t(\T)\pi(d\t)$ is approximated by
$$
\frac{1}{M}\sum_{i=1}^M
\sigma_{m-n_{(i)}-1}(\S^-_{(i)}|\theta_{(i)})\,
\sigma_{n_{(i)}-1}(\S^+_{(i)}|\theta_{(i)}) \,
\rho(\theta_{(i)}),
$$
if $(\S_{(1)}^-,\S_{(1)}^+,\t_{(1)}),\ldots,
(\S_{(M)}^-,\S_{(M)}^+,\t_{(M)})$ are independent samples
obtained via implementing the SIP$(\t)$ sampler, whereby
$n_{(i)}$ is determined in step~(S1) after $\t_{(i)}$ is
fixed in step~(S0), and
$\sigma_{m-n_{(i)}-1}(\S^-_{(i)}|\theta_{(i)})$ and
$\sigma_{n_{(i)}-1}(\S^+_{(i)}|\theta_{(i)})$ are obtained
from steps~(S2) and~(S3), respectively.

\section{Proportional Hazards}\label{sec:cox}
The Cox regression model~[Cox~(1972)] is an important
example of the multiplicative intensity model that can
allow incorporation of covariates, together with right
independent censoring, in survival analysis. For Bayes
inference of general hazard rates with presence of
covariates, see Kalbfleisch~(1978), Ibrahim, Chen and
MacEachern~(1999), James~(2003) and Ishwaran and
James~(2004), among others. Suppose we collect failure data
until time $\tau$, which are governed by an underlying
hazard rate on $\Rp$ associated with a $p$-dimensional
covariate vector $\X\in {\R}^p$,
$$
\lambda(t|\X, \Bt, \mu, \t) =
\lambda(t|\mu,\t)\exp({\Bt}^{T}\X),
$$
where $\lambda(t|\mu,\t)$ defined in~(\ref{bathtub}) is an
unknown baseline hazard rate of a bathtub shape and
$\Bt\in{\R}^p$ is an unknown parameter vector. The data
$\mathbf{D} = ((T_1,\X_1), \ldots, (T_N,\X_N))$ summarize
completely observed failure times $T_1 < \cdots < T_m$ and
right-censored times $T_i=\tau$, $i=m+1,\ldots,N$,
associated with covariate vector $\X_i$, $i=1,\ldots,N$,
respectively. Define $\fNt(x,u) = \gNt(u)x$, for any $(x,u)
\in (\Rp, \R)$, where
\begin{equation}\label{gN}
    \gNt(u) = \int_{0}^{\tau} \left[\sum_{i=1}^{N} \I(T_i \geq t)
    \exp({\Bt}^{T}\X_i)\right] [\I(t-\t \leq u < 0) + \I(0 < u \leq t-\t)] dt.
\end{equation}
Then, the Cox proportional hazards likelihood may be
written as
\begin{equation}
\label{like2}
    \left[ \prod_{i=1}^{m}\exp({\Bt}^{T}\X_i)\lambda(T_i|\mu,\t) \right]
    \exp \left[ -\mugNt\right],
\end{equation}
where $\mugNt = \int_{\R} \gNt(u) \mu(du) = \int_0^{\tau}
[\sum_{i=1}^{N} \I(T_i \geq t) \exp({\Bt}^{T}\X_i)]
\lambda(t|\mu,\t) dt$. Assume\break $\int_{\R} x^{\ell}
\e^{-\gtNt(u)x} \rho(dx|u) < \infty$, for $\ell=1,\ldots,m$
and a fixed $u>0$. If $\pi(d\Bt)$ and $\pi(d\t)$ are
independent priors for $\Bt$ and $\t$, applying the same
arguments in proving Theorems~\ref{Prop1} and~\ref{combin}
yields that the law of $\mu|\mathbf{D}$ is equivalent to
that of a random measure $\mu_{\gtNt} +
    \sum_{\{j^\ast|\S^-\}} Q_j^- \delta_{y_j^-}+
    \sum_{\{j^\ast|\S^+\}} Q_j^+ \delta_{y_j^+}$, where $\mu_{\gtNt}$, with law denoted by
$\P(d\mu_{\gtNt})$, is a completely random measure with
L\'evy measure $\e^{-\gtNt(u)x} \rho(dx|u) \eta(du)$. It is
determined by the law of\break $\mu_{\gtNt},
\Q^-,\y^-,\S^-,\Q^+,\y^+,\S^+,\t,\Bt|\mathbf{D}$, which is
proportional to
\begin{eqnarray*}
    &&\hspace*{-0.5in}\P(d\mu_{\gtNt}) \pi(d\t) \pi(d\Bt)
    {\cal L}_{\mu}(\gNt|\rho, \eta) \prod_{i=1}^{m}
    \exp({\Bt}^{T}\X_i)\nonumber\\
    &&\hspace*{-0.2in}
    \times|{\C}_{\S^-}|\prod_{\{j^{\ast}|\S^-\}} \left\{({Q_j^-})^{m_j^-}
    \e^{-\gtNtt(y_j^-) Q_j^-} \rho(dQ_j^- | y_j^-)
    \I(Z_j^\t \leq  y_j^- < 0)
    \eta(dy_j^-)\right\}\nonumber \\
    &&\hspace*{-0.2in}  \times
    |{\C}_{\S^+}|\prod_{\{j^{\ast}|\S^+\}} \left\{({Q_j^+})^{m_j^+}
    \e^{-\gtNtt(y_j^+) Q_j^+} \rho(dQ_j^+ | y_j^+)
    \I(0 < y_j^+ \leq Y_j^\t) \eta(dy_j^+)\right\}.
    \label{jointSQ2}
\end{eqnarray*}
Analogous results with presence of covariates of Theorems~\ref{Prop1} and~\ref{combin}
in terms of two $\S$-paths can be obtained via Bayes'
theorem and multiplication rule.

\begin{prop}\label{Prop2}{\rm
Suppose the likelihood of the data is given
by~(\ref{like2}). Assume that $\mu$ is a completely random
measure characterized by the Laplace
functional~(\ref{Lap}), and independently, let $\pi(d\Bt)$
and $\pi(d\t)$ denote independent priors for $\Bt$ and
$\t$. Then,
\begin{enumerate}
    \item[(i)] the law of $\mu|\t,\Bt,\mathbf{D}$ can be described by a three-step
hierarchical experiment as in Theorem~\ref{Prop1}, of which
$\fNtt(\cdot,\cdot)$ and $\gNtt(\cdot)$ are replaced by
$\fNt(\cdot,\cdot)$ and $\gNt(\cdot)$, respectively.
    \item[(ii)] the law of $\t|\Bt,\mathbf{D}$ is
characterized by, for any Borel set $B \in \Rp$,
$$
  \Pr(\theta \in B|\Bt,\mathbf{D}) = \int_B \sum_{\S^-} \sum_{\S^+}
  \pi(\S^-,\S^+,d\theta|\Bt,\mathbf{D})\label{posttheta2},
$$
where $\pi(\S^-,\S^+,d\theta|\Bt,\mathbf{D}) \propto {\cal
L}_{\mu}(\gNt|\rho, \eta)
    \times |\mathbb{C}_{\S^-}| \prod_{\{j^{\ast}|\S^-\}}  \int_{Z^\t_j}^{0}
    \kappa_{m_j^-}(\e^{-\ftNt} \rho|y)
    \eta(dy)\break\times
    |\mathbb{C}_{\S^+}| \prod_{\{j^{\ast}|\S^+\}} \int_{0}^{Y^\t_j}
    \kappa_{m_j^+}(\e^{-\ftNt} \rho|y)
    \eta(dy)\times \pi(d\t)$.
\end{enumerate}    }
\end{prop}

To evaluate any posterior quantities of
model~(\ref{like2}), such as the posterior mean of the
underlying bathtub-shaped baseline hazard rate and the
posterior mean of the covariate parameters $\Bt$, run the
following Gibbs sampler to obtain random samples from the
posterior distribution of
$(\Q^-,\y^-,\S^-,\Q^+,\y^+,\S^+,\t,\Bt)$ given
$\mathbf{D}$:
\begin{enumerate}
    \item Draw $\S^-,\S^+|\Q^-,\y^-,\Q^+,\y^+,\S^+,\t,\Bt,\mathbf{D}$ by independently
    implementing Algorithm~\ref{AP} as in steps~(M1) and~(M2) in Section~\ref{sec:MCMC}.
    \item Draw $\Q^-,\y^-,\Q^+,\y^+|\S^-,\S^+,\t,\Bt,\mathbf{D}$ according to the analogues of
    the conditional distributions~(\ref{alphaSjn}--\ref{Qjp})
    in Theorem~\ref{Prop1} with $\fNtt(\cdot,\cdot)$ and $\gNtt(\cdot)$ replaced by
    $\fNt(\cdot,\cdot)$ and $\gNt(\cdot)$, respectively.
    \item Draw $\t|\Q^-,\y^-,\S^-,\Q^+,\y^+,\S^+,\Bt,\mathbf{D}$ from the density proportional to
    $$\pi(d\t)  {\cal L}_{\mu}(\gNt|\rho, \eta) \prod_{\{j^{\ast}|\S^-\}} \e^{-\gtNt(y_j^-) Q_j^-}\I(Z^\t_j-\t \leq y_j^-)
    \prod_{\{j^{\ast}|\S^+\}}\e^{-\gtNt(y_j^+) Q_j^+}\I(y^+_j \leq Y^\t_j-\t).$$
    \item Draw $\Bt|\Q^-,\y^-,\S^-,\Q^+,\y^+,\S^+,\t,\mathbf{D}$ from the density proportional to
    $$\pi(d\Bt)  {\cal L}_{\mu}(\gNt|\rho, \eta) \prod_{i=1}^{m} \exp({\Bt}^{T}\X_i)
    \prod_{\{j^{\ast}|\S^-\}} \e^{-\gtNt(y_j^-) Q_j^-}
    \prod_{\{j^{\ast}|\S^+\}}\e^{-\gtNt(y_j^+) Q_j^+}.$$
\end{enumerate}

Note that $\gNt(u)$ is again a piecewise linear function of
$u$ as $\gNtt(u)$ in the case without covariates. This does
not create any complexities in evaluating integrals at
steps~1 and~2 of the above Gibbs sampler~(see discussion of
Remark 5.1 in Ho~(2006a)). Step~4 above, which is of the
same form as the step 4~(for conditional draws of
regression parameters $\Bt$) of the Blocked Gibbs algorithm
suggested by Ishwaran and James~(2004, page 184), can be
dealt with via a Metropolis step, while step~3 can also be
done similarly as the density looks like the one in step~4.





\end{document}